\documentclass[pdflatex,sn-standardnature]{sn-jnl}

\jyear{2026}%

\theoremstyle{thmstyleone}%
\theoremstyle{thmstyletwo}%
\usepackage{microtype}

\theoremstyle{thmstylethree}%

\usepackage{lineno}

\usepackage{mhchem} 
\usepackage{fancyhdr}
\usepackage{amsmath}

\usepackage{makecell}
\usepackage{booktabs}
\usepackage{amssymb}
\usepackage{graphicx} 
\usepackage{amsmath}
\numberwithin{equation}{section} 

\usepackage[T1]{fontenc}

\begin{document}

\title{Direct experimental measurement of femtonewton-scale momentum transfer force from electron beams}


\author[1]{\fnm{Chunbo} \sur{Lin}}

\author*[2,3]{\fnm{Xinggang} \sur{Shang}}\email{shangxinggang@westlake.edu.cn}

\author[1,4,5]{\fnm{Xijun} \sur{Li}}

\author[1]{\fnm{Xiaoyu} \sur{Sun}}

\author[5]{\fnm{Kang} \sur{Zhao}}

\author[1]{\fnm{Xujie} \sur{Wang}}

\author[1]{\fnm{Yang} \sur{Yu}}

\author[6]{\fnm{Wenjing} \sur{Cao}}

\author*[1,5]{\fnm{Min} \sur{Qiu}}\email{qiu\_lab@westlake.edu.cn}

\affil[1]{Zhejiang Key Laboratory of 3D Micro/Nano Fabrication and Characterization, Department of Electronic and Information Engineering, School of Engineering, Westlake University, Hangzhou, Zhejiang 310030, China}

\affil[2]{Department of Mechanical and Automation Engineering, The Chinese University of Hong Kong, Shatin, Hong Kong, China}

\affil[3]{Cerebro-Cardiovascular Health Engineering (COCHE), Shatin, N.T., Hong Kong, China}

\affil[4]{Center for Micro/Nano Fabrication, Westlake University, Hangzhou, Zhejiang 310030, China}

\affil[5]{Westlake Institute for Optoelectronics, Fuyang, Hangzhou, Zhejiang 311400, China}

\affil[6]{Instrumentation and Service Center for Physical Sciences, Westlake University, Hangzhou, Zhejiang 310030, China}


\abstract{

Electron beams (e-beams) are ubiquitous in imaging, patterning, and propulsion. This prevalence is rooted in the profound mastery of their wave-particle duality and energy-transfer pathways. Yet, a fundamental dimension remains largely unexplored: while the mechanical effect (i.e., the momentum transfer to a target) is theoretically known, quantification of its femtonewton-range force has remained elusive. This discrepancy represents a missing piece of the puzzle toward a comprehensive understanding of e-beam–matter interactions, and ultimately limits the multi-dimensional exploitation of e-beams. A force sensor combining femtonewton sensitivity, immunity to electromagnetic noise, compatibility with vacuum, and absolute calibration is critical to bridge the gap between theory and experiment. Here the FINEST (Femtonewton Interferometric Nanomechanical Electron-beam Sensing Technology) sensor is proposed and successfully tested to measure the force of an e-beam. FINEST is an optical-pressure-calibrated 3D spring-type optical sensor that operates reliably under e-beam conditions. Femtonewton‑scale forces from 2–30 keV e-beams are directly measured, ranging from 505 fN to 13 pN. Both linear scaling with beam current and a non‑monotonic energy dependence (peaking near 10 keV) are observed. Based on this calibrated force, the mechanical contribution to e-beam ice etching was quantitatively confirmed; its effect is orders of magnitude lower than the total etch depth and lacks noticeable energy dependence. By achieving the first direct experimental measurement of e-beam momentum transfer, this work adds a long-missing dimension to the physical landscape of e-beam processes. These findings provide a quantitative basis for furthering the multi-dimensional exploitation of e-beams, potentially transforming our approach to precision nanofabrication, sensing, and fundamental electron physics research.}




\maketitle

\thispagestyle{plain}
\pagestyle{plain}

\section{Introduction}
The electron beam is a cornerstone of modern science and industry, with various applications in macroscopic phenomena, such as charged particle fluxes interacting with satellites in complex plasma environments \cite{obayashi1984space}, high-energy particle streams driving free-electron lasers \cite{deacon1977first} and advanced accelerators \cite{litos2014high}, precision tools for high-end cutting and welding \cite{wkeglowski2016electron} and in the microscopic realm, such as SEM \cite{von1938elektronen}, electron-beam-induced deposition \cite{huth2012focused}, and electron-beam lithography \cite{broers1976electron}. These applications explore electron–matter interactions based on energy deposition or electron-molecular interactions while a fundamental physical picture, namely mechanical force, has remained largely overlooked. 

In a picture of electron-beam interaction with materials, the inherent momentum of the beam may exert a mechanical force on the target material. However, due to various constraints, this momentum transfer has not been quantified and has been largely overlooked, which obscures novel perspectives for exploring mechanisms in structural biology, astrophysics, micro-nanomanufacturing, and beyond. Theoretically, the Monte Carlo method integrated with energy deposition, for example, the CASINO software \cite{drouin2007casino}, can simulate this transfer process. However, it cannot accurately model the vector momentum flux involved in this physical process. Furthermore, the microscopic mechanism of electron momentum partition between elastic recoil and inelastic excitation remains not fully understood \cite{susi2019quantifying}, inherently limiting the precise calculation of macroscopic descriptors such as the dimensionless correction factor ($\eta$) derived from the scattered electron spectra. Experimentally, precise capture of the exceptionally weak mechanical signals of an e-beam is more challenging. To date, only qualitative and indirect momentum transfer from e-beam to nanowires was reported \cite{pairis2019shot}, and the specific role of the electron-beam-induced force in nanoscale stress distribution, structural deformation, and e-beam fabrication precision needs detailed and quantitative investigation. The primary obstacle to this is the lack of an accurate measuring methodology for electron-beam-induced mechanical forces. Conventional micro- and nanomechanical force sensors, such as piezoresistive sensors \cite{fiorillo2018theory}, capacitive microelectromechanical systems \cite{mishra2021recent}, and microcantilevers with electrical readouts \cite{vashist2007review}, rely on closed-loop electrical circuits which are fragile to high-energy electrons, the accumulated surface charge, and interfered by intense spatial electromagnetic fields. Furthermore, local thermal drift induced by continuous electron bombardment further deteriorates the baseline stability of active electrical sensors. Consequently, these characteristics of electrical detection mechanisms render these sensors incapable of the detection sensitivity and interference resistance required by measuring the force of an e-beam. All-optical sensing mechanisms have gradually emerged to overcome the bottlenecks of conventional sensors, owing to their exceptional immunity to electromagnetic interference and ultrahigh displacement resolution \cite{liu2020integrated}. However, directly capturing the momentum/force signal of e-beams imposes stringent constraints on the measurement platform. An ideal system must possess in situ measurement capabilities within the electron-beam apparatus, ultrahigh device integration, and strict environmental compatibility with vacuum. Most existing optical measurement techniques are fundamentally incompatible with these requirements. For instance, atomic force microscopy (AFM) \cite{binnig1986atomic} features a bulky optical path that blocks its full integration into the chamber of a scanning electron microscope. Optical tweezer technology \cite{jones2015optical} depends on liquid suspension environments or specific atmospheric refractive indices and does not work in the vacuum conditions of an operating electron-beam. Nanosensors integrated on optical fiber facets \cite{li2020ultrathin} can work in the chambers of electron microscopes, but existing fiber facet sensors can detect force ranging only from piconewton (pN) to micronewton \cite{zou2021fiber, li2025fish}. Specifically, to achieve ultimate mechanical sensitivity, an ultra-slender or ultrathin structure is needed, which lacks a sufficient interaction volume for a high-energy e-beam, and misses at least part of the e-beam momentum transfer (see Extended Data Fig.~\ref{Ex0}). 

A three-dimensional micro-helical architecture offers an elegant solution to innovate the mechanical sensor to detect the momentum transfer of an e-beam to materials. It not only compactly folds an ultralong effective lever arm to ensure excellent overall stability, but also provides a broad and flat e-beam receiving area for uniform stress distribution. More crucially, the intrinsic structural properties of the helical spring dictate that increasing the helical radius to expand the electron receiving area does not degrade device performance; it rather drastically reduces the system stiffness by a cubic factor ($k \propto 1/R^3$) \cite{shang2024fiber}, thereby significantly enhancing the probe sensitivity. Unlike cantilevers that are highly susceptible to nonlinear bending, this helical spring architecture maintains a standard and stable linear response over a substantially large deformation range. Building upon these, this work reports the Femtonewton Interferometric Nanomechanical Electron-beam Sensing Technology (FINEST) to detect the force of an e-beam. By integrating an optical-pressure-calibrated, three-dimensionally printed spring-suspended optical probe with an interferometric system, our sensor resolves the long-standing challenge of experimentally quantifying electron-beam momentum and pressure. This methodology features the distinct advantages of femtonewton-level force detection, electromagnetic interference immunity, and ultracompact size for in situ integration, making it most compatible with e-beam conditions. The core of FINEST is a three-dimensional micro-helical optomechanical coupling element of an exceptionally low spring constant of 0.00078 N/m and an ultrahigh sensitivity of 13 pm/pN. It is fabricated utilizing multiple fabrication techniques, such as two-photon three-dimensional printing and focused ion beam milling. Following rigorous calibration via optical radiation pressure, FINEST is applied to measure the forces of an e-beam accelerated from 2 to 30 keV, with measured forces ranging from 505 fN to 13 pN. The experimental results quantitatively delineate the exerted force scaling linearly with the e-beam current, and further reveal a non-monotonic energy dependence, thereby advancing the physical understanding of electron–matter interactions. Furthermore, the experimental findings are applied to investigate the physics of electron-beam lithography of ice, which can also be explored to elucidate the mechanisms of nanomaterial manipulation by e-beams. These discoveries provide a solid quantitative foundation for a deep understanding of the mechanics of the interaction of e-beams with materials and practical guidance for process development and innovation of micro- and nanofabrication, as well as micro/nano manipulation with e-beams.

\section{Principle and realization of the FINEST}

Achieving precise measurement of e-beam pressure and momentum transfer currently faces three main challenges: (i) electromagnetic immunity, necessitating resistance to interference within strong electromagnetic environments; (ii) in-situ integration, requiring the incorporation of detection devices into vacuum electron chambers; and (iii) femtonewton-scale force sensing, demanding suitable methodologies for resolving weak forces at the $\mathrm{fN}$ level. The FINEST platform effectively resolves these three challenges simultaneously.

\begin{figure}[htbp]
\centering
\includegraphics[width=1\textwidth]{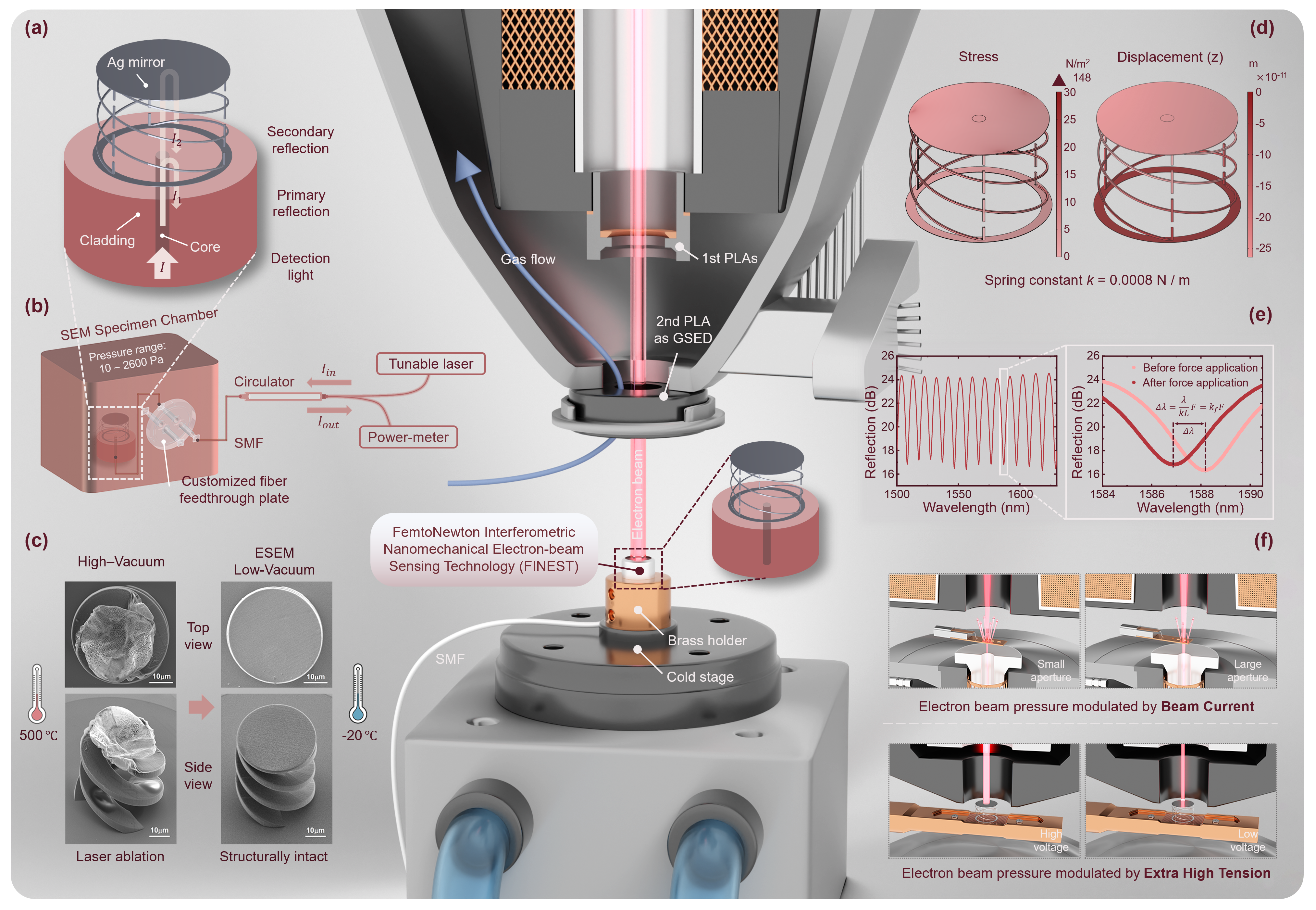}
\caption{\textbf{Femtonewton Interferometric Nanomechanical Electron-beam Sensing Technology (FINEST).} \textbf{(a)} Structure and operating principle of FINEST. A 3D helical micro-spring is fabricated on the end face of an optical fiber by a hybrid additive--subtractive process. Light from a tunable laser is reflected by the fiber end face and the top plate of the helical structure, generating two interfering beams, $I_1$ and $I_2$, that define a fiber-tip Fabry--P\'{e}rot cavity. \textbf{(b)} Experimental setup for electron-beam force sensing. The probe is positioned inside a temperature- and pressure-controlled SEM chamber, with the optical fiber routed out through a vacuum feedthrough. A tunable laser and an optical power meter are connected to the probe via an optical circulator outside the chamber. Centre, enlarged view of the in-chamber probe configuration. \textbf{(c)} Environmental scanning electron microscopy enables the probe to remain structurally intact and operational under simultaneous laser illumination and electron-beam irradiation. \textbf{(d)} Simulated stress and displacement distributions of the probe under a 1~pN vertical load applied to the top plate. \textbf{(e)} Reflection spectrum as a function of laser wavelength, with a zoomed-in view of the resonance minimum. Force readout is based on the Fabry--P\'{e}rot resonance shift. \textbf{(f)} Schematic of electron-beam pressure modulation by beam current and extra high tension (EHT).}
\label{Fig1}
\end{figure}

Reliable device operation under e-beam bombardment is the fundamental prerequisite for successfully measuring e-beam pressure. The FINEST is a measurement platform based on interferometric principles. It eliminates electromagnetic interference through an all-optical closed-loop link. Simultaneously, the metallized shell of the FINEST establishes an excellent active charge dissipation channel. Serving as the core of the FINEST, the FINEST probe (illustrated in Fig.~\ref{Fig1}a) features a classic Fabry-Pérot (F-P) cavity structure fabricated on the facet of a single-mode fiber (SMF), where applied forces induce changes in the cavity length. As shown in Fig.~\ref{Fig1}b, a tunable laser outside the electron chamber emits a beam for spectral sweeping. The laser passes through a circulator and is subsequently introduced to the probe inside the chamber via a customized fiber feedthrough plate. The laser undergoes two reflections at the fiber facet and the top plate of the ultrafast spring. This generates two distinct beams, $I_1$ and $I_2$. These two beams subsequently interfere with each other. The interference signal passes back through the circulator and is received by a power meter. The mechanical displacement of the FINEST probe is converted into a selective shift of the intrinsic wavelength of the resonant cavity. This wavelength modulation enables remote spectral demodulation and precise mechanical quantification. The entire process relies solely on changes in the resonance spectrum and remains completely free from electrical interference.

Integrating the device in-situ into the electron chamber and ensuring its stable operation in a vacuum environment are essential conditions for measuring e-beam pressure. The FINEST probe is only $200~\mathrm{\mu m}$ in size. Furthermore, the integrated fiber facet is less than $2~\mathrm{cm}$ in height. The interference signal is directly transmitted out of the chamber through a slender and flexible optical fiber. These compact dimensions perfectly accommodate the conventionally narrow SEM e-beam chamber. Furthermore, active optical devices often face severe thermal challenges when operating in a vacuum environment. The thermal conductivity of the FINEST probe was optimized. Alongside this optimization, both ESEM technology and active cooling were implemented (see Fig.~\ref{Fig1}c and \textcolor{blue}{Methods \ref{8.1}}). To ensure structural stability and maintain a low operating temperature, the FINEST probe and the fiber are mounted on a highly thermally conductive brass holder. This holder is directly connected to a cold stage. This assembly is mounted as a single unit on a 5-axis motorized stage within the ESEM system. It is positioned directly beneath the electron column (see the middle panel of Fig.~\ref{Fig1}).

Traditional mechanical elements, such as cantilevers, are fundamentally constrained by the trade-off between sensing volume and measurement precision. A spring, a ubiquitous structure in everyday applications, resolves this contradiction, successfully enabling femtonewton ($\mathrm{fN}$)-scale measurements over a force-receiving area on the order of hundreds of micrometers. The employed helical spring is a common type, comprising an annular base, a triple-start helix, and a circular top plate. As the primary sensing component of the FINEST platform, the probe represents a micro- and nanoscale evolution of this classic architecture. To our knowledge, this marks the first application of such a structure in the field of particle mechanics detection. This spring design is extremely sensitive to external forces. Furthermore, its deformation exhibits a linear response within a remarkably large elastic regime. Simulations were conducted to evaluate the stress distribution and the displacement field along the $z$-axis of the probe under a $1~\mathrm{pN}$ external force applied to the center of the top plate (see \textcolor{blue}{Methods \ref{8.2}}). Subsequently, its spring constant ($k$) was experimentally calibrated to be $0.0008~\mathrm{N~m^{-1}}$ (Fig.~\ref{Fig1}d).

This optomechanical coupling structure converts the momentum signal of the e-beam into an optical signal. Fig.~\ref{Fig1}e (left) illustrates the interference spectrum acquired by the interferometry system when the probe is free from external forces. E-beam bombardment on the circular top plate induces a structural deformation in the helical probe. Consequently, the length of the F-P cavity formed by the top plate and the SMF facet undergoes a nanometer-scale change. This subtle variation is captured by the sweeping tunable laser. It manifests as a global shift in the spectral resonance, as shown in the right panel of Fig.~\ref{Fig1}e. By measuring the shift of the spectral peaks or valleys, the magnitude of the force exerted by the e-beam on the probe can be accurately determined according to the equation $F = \frac{kL}{\lambda}\Delta\lambda = \frac{\Delta\lambda}{k_f}$.

Collectively, the FINEST platform comprises three main components. These are a probe, a modified environmental scanning electron microscope (ESEM) system, and an interferometry system. Specifically, the probe detects the e-beam pressure. The ESEM provides the working environment and emits the e-beam. Finally, the interferometry system acquires the mechanical signals and converts them into optical signals.

Baseline robustness and sensitivity tests (see Extended Data Fig.~\ref{Ex5} and \textcolor{blue}{Supplementary Note 1}) demonstrate that the theoretical measurement limit of the FINEST fully meets the sensitivity requirements for fN-scale e-beam pressure measurements. Building upon this capability, experiments reveal that the force exerted by the e-beam can be modulated by varying the beam current and the extra high tension (EHT) (Fig.~\ref{Fig1}f). Consequently, we systematically investigated the specific force values, scaling trends, and underlying physical mechanisms across different e-beam parameters. This approach yields a wealth of experimental data previously inaccessible through either experiment or simulation. Furthermore, it is accompanied by comprehensive theoretical interpretations.

\section{Architecturally inspired micro-fabrication}

\begin{figure}[htbp]
\centering
\includegraphics[width=1\textwidth]{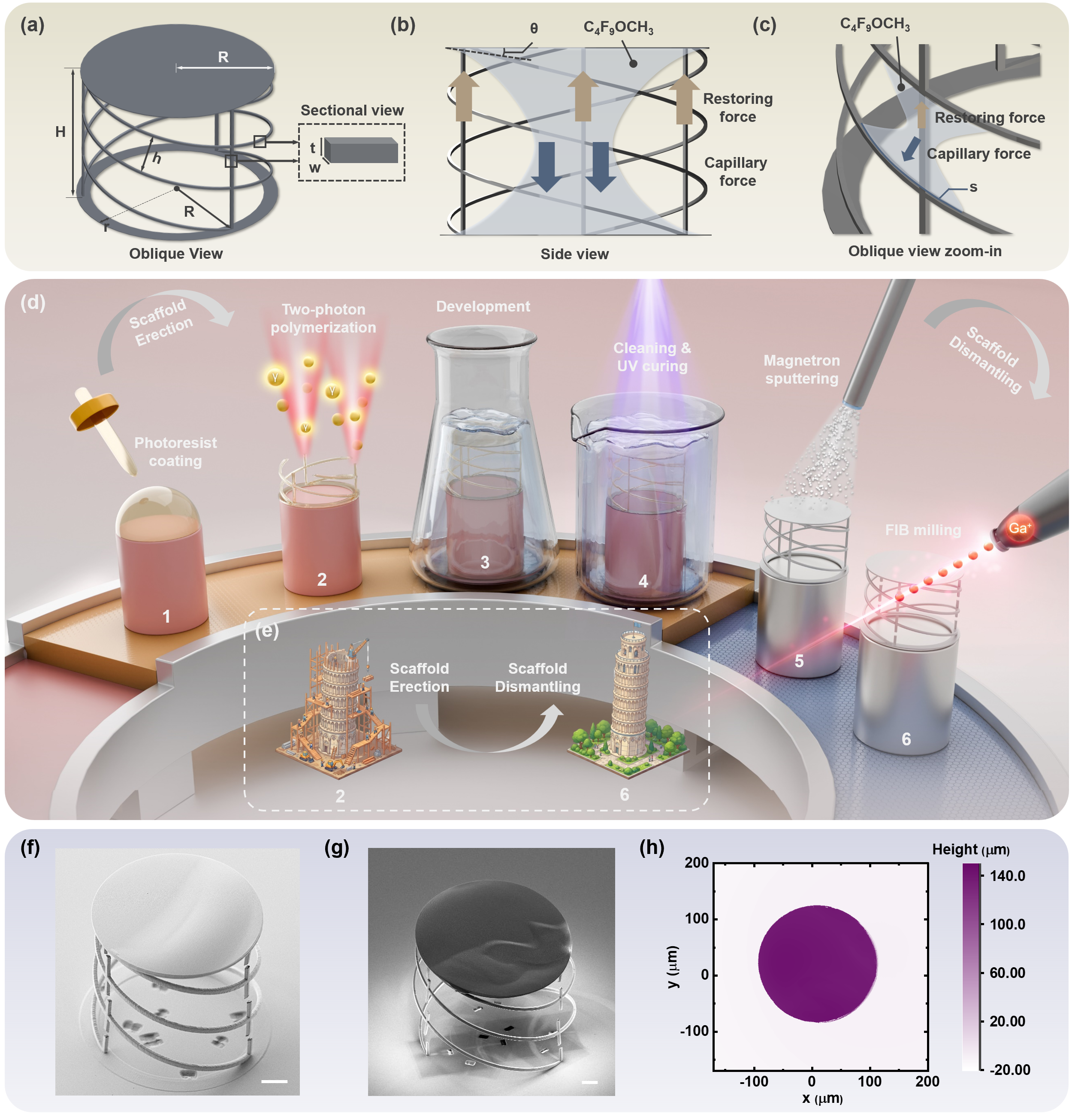}
\caption{\textbf{Fabrication and structural characterization of the FINEST probe.} \textbf{(a)} Oblique-view schematic defining the probe geometry. $H$, height from the top plate to the base plane; $h$, spacing between adjacent helical turns; $R$, helix radius; $r$, width of the bottom annular ring; $t$ and $w$, cross-sectional dimensions of the helical beam, as defined in the sectional view. \textbf{(b, c)} Side-view and zoomed-in oblique-view schematics illustrating the mechanical role of the vertical scaffold beams. During solvent cleaning, capillary forces from the liquid act on the probe, whereas the three scaffold beams provide a restoring force that preserves structural integrity. \textbf{(d)} Hybrid additive--subtractive fabrication workflow of the FINEST probe. Steps 1--4 involve TPL fabrication of the probe, followed by deposition of an $\sim$25~nm Ag layer by high-vacuum magnetron sputtering (step 5) and Ga$^+$ FIB milling (step 6). \textbf{(e)} Schematic of scaffold erection and dismantling in the FINEST fabrication strategy. Inspired by architectural construction, the helical framework is first erected and the vertical support structures are subsequently removed. \textbf{(f)} SEM image of a FINEST probe with $R = 65~\mu$m. Scale bar, 20~$\mu$m. \textbf{(g)} FIB image of a FINEST probe with $R = 100~\mu$m. Scale bar, 20~$\mu$m. \textbf{(h)} Laser confocal scanning image of a FINEST probe with $R = 100~\mu$m, showing the lateral dimensions and height profile.}
\label{Fig2}
\end{figure}

The FINEST probe features a typical complex three-dimensional freestanding architecture, presenting an inherent tradeoff between the detection limit and fabrication complexity. Theoretically, the spring constant $k$ of helical coils with a rectangular cross-section can be described by
\begin{linenomath*}
\begin{equation}
k = \frac{G t^2 w^2}{\gamma n (2R)^3},
\end{equation}
\end{linenomath*}
where $G$ is the shear modulus of the material, $\gamma$ is a coefficient dependent on $w/t$, and $n$ is the number of coil turns\cite{shang2024fiber}. Governed by this underlying physical relationship, a probe with a larger $R$ and smaller $t$, $w$, and $H$ (Fig.~\ref{Fig2}a) effectively reduces the system stiffness, thereby exhibiting higher sensitivity. However, it is considerably more challenging to fabricate. Conventional monolithic two-photon polymerization techniques cannot realize such highly demanding structures.


Specifically, the most significant challenge encountered throughout the fabrication process is the structural damage caused by capillary forces\cite{shang2022customizable, delrio2005role} during liquid-phase processing steps (Fig.~\ref{Fig2}b, c). Minute forces can induce large displacements in the helical structure. This characteristic ensures extreme sensitivity. However, it inherently causes low structural stiffness and a high susceptibility to deformation during fabrication. During this procedure, the probe must detach from the liquid surface twice following the completion of step 3 and step 4 (Fig.~\ref{Fig2}d). Upon detachment, residual liquid on the probe surface forms liquid bridges. These bridges span the gap between the top plate and the fiber facet (Fig.~\ref{Fig2}b). They also form between adjacent helical coils (Fig.~\ref{Fig2}c). During evaporation, these liquid bridges exert strong capillary forces directed along the blue arrows. This tension typically causes the probe to collapse (see Extended Data Fig.~\ref{Ex1}a).

This difficulty was resolved by drawing inspiration from architectural scaffolding structures. In macroscopic construction, scaffolding serves as a temporary structure to support and shape the main building\cite{hunt1981review}. It is subsequently removed once the primary structure achieves stability (Fig.~\ref{Fig2}e). Before structures like long-span beams, arches, or complex concrete pours acquire sufficient self-supporting capabilities, the scaffold bears the majority of the gravitational load. This concept was adapted to the fabrication of the FINEST probe. During the initial two-photon polymerization stage, three rectangular pillar supports connected to the helix are fabricated (Fig.~\ref{Fig2}b, c). These three scaffold beams counteract the surface tension by providing a restoring force (yellow arrows). This support mechanism ensures the probe maintains its structural integrity and stability throughout the fabrication process. These three pillars are subsequently removed using FIB milling in the final microfabrication step. The assembly and removal of the scaffolding correspond to fabrication step 2 and step 6 (Fig.~\ref{Fig2}d), respectively. An established capillary force model provides further theoretical validation for this mechanism (see the \textcolor{blue}{Supplementary Note 2}).

Fig.~\ref{Fig2}d illustrates the fabrication workflow of the FINEST probe. This workflow comprises six distinct steps. These steps include photoresist coating, two-photon polymerization, development, cleaning and UV curing, magnetron sputtering, and FIB milling. Optical and electron micrographs corresponding to steps 1 through 6 are presented in Extended Data Fig.~\ref{Ex1}b-e. Furthermore, intuitive schematics of step 2 and step 6 are displayed in Fig.~\ref{Fig2}e. This approach integrates additive manufacturing (TPL and deposition in steps 1 through 5) with subtractive manufacturing (FIB milling in step 6). Furthermore, six specific optimizations were implemented across the entire workflow from design to fabrication (see \textcolor{blue}{Methods \ref{8.3}} and Extended Data Fig.~\ref{Ex1}). This combined process represents a newly developed strategy for fabricating helical structures with extremely small spring constants ($k$). It has been proven highly effective during actual fabrication.

Two fabricated probes are presented with baseline parameters of $t = 2.5~\mathrm{\mu m}$ and $w = 2.5~\mathrm{\mu m}$. They feature distinct radii and heights of $R = 65~\mathrm{\mu m}$ and $H = 115~\mathrm{\mu m}$ (Fig.~\ref{Fig2}f, denoted as R65), and $R = 100~\mathrm{\mu m}$ and $H = 150~\mathrm{\mu m}$ (Fig.~\ref{Fig2}g, denoted as R100). The R65 design demonstrates a near-100\% fabrication yield. The R100 design represents the highest sensitivity achievable with current fabrication capabilities. Confocal microscopy was utilized for the morphological characterization of R100 (Fig.~\ref{Fig2}h, Extended Data Fig.~\ref{Ex6} and \textcolor{blue}{Supplementary Note 3}). The fabricated dimensions exhibit high fidelity to the target design. Furthermore, the surface roughness of the top plate is well-controlled at an optimal level of several hundred nanometers. This specific probe design is utilized for all subsequent calibrations and e-beam pressure measurements.

\section{Force calibration via photon momentum}

Accurate measurement of e-beam pressure requires a probe calibrated with absolute force values. Traditional mechanical calibration methods often rely on the gravitational conversion of microscale particles or stiffness derivations of cantilevers. However, at the femtonewton (fN) microscopic scale, these approaches are time-consuming and suffer from limited precision. For example, calculating particle weight relies on estimations of volume and density\cite{martinez2017inertial}. Furthermore, physical contact during calibration introduces additional damage to the probe. To overcome this limitation, we introduced a non-contact calibration scheme based on photon momentum transfer. Radiation pressure is generated by illuminating the micro-helical structure with a laser. By precisely controlling the laser power and wavelength, the theoretical absolute thrust can be calculated. This calibration method is directly founded on fundamental physical constants and has been experimentally verified as reliable\cite{yu2024optical, zhu2025gold}.

\begin{figure}[htbp]
\centering
\includegraphics[width=0.81\textwidth]{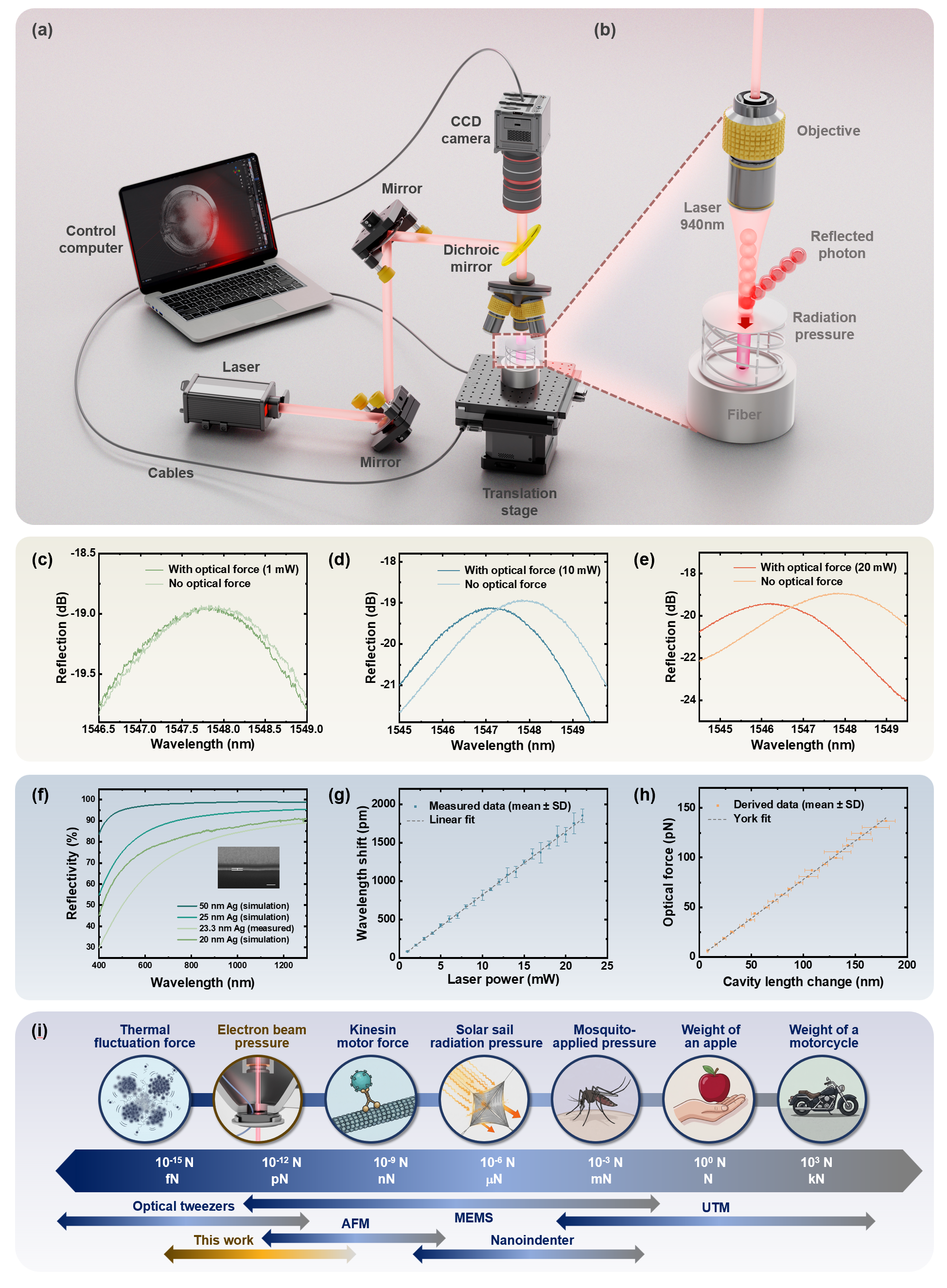}
\caption{\textbf{Optical radiation-pressure calibration of the FINEST probe.} \textbf{(a)} Optical calibration platform for the FINEST probe. A 940~nm laser is directed onto the probe through two Ag mirrors, a short-pass dichroic mirror, and a microscope objective, generating radiation pressure on the top plate. The CCD camera provides real-time imaging of the probe under coaxial white-light illumination, enabling the laser to be positioned at the centre of the helical spring. \textbf{(b)} Principle of optical radiation-pressure generation. Momentum transfer from the incident and reflected photons exerts a normal force on the top plate, deforming the helical spring. \textbf{(c--e)} Interference resonance spectra of the FINEST probe measured at optical powers of 1, 10 and 20~mW, respectively. Dark curves denote spectra recorded under laser illumination, and light curves denote those recorded without optical loading, revealing progressive resonance shifts with increasing optical power. \textbf{(f)} Reflectivity spectrum of the top plate, obtained from microspectrophotometer measurements and finite-difference time-domain simulations. Inset, cross-sectional SEM image of the processed top plate, with the Ag-coating thickness marked. Scale bar, 100~nm. \textbf{(g)} Resonance-wavelength shift as a function of laser power. Error bars indicate s.d., and the grey dashed line shows a linear fit to the measured data. \textbf{(h)} Optical force as a function of Fabry--P\'{e}rot cavity-length change. Error bars represent s.d.\ propagated from the resonance-shift uncertainty, and the grey curve shows a York fit. The slope gives the spring constant $k$ of the FINEST probe. \textbf{(i)} Comparison of force scales and representative force-characterization techniques, highlighting the operating regime of the present work.}
\label{Fig3}
\end{figure}

Fig.~\ref{Fig3}a illustrates the optical experimental setup for this calibration. The FINEST probe is mounted on a fiber pedestal. It is firmly attached to a three-axis motorized translation stage to ensure stability. A $940\text{-}\mathrm{nm}$ single-frequency laser generates the laser beam for calibration. The translation stage is moved to focus the laser onto the center of the top plate of the probe. This generates a Gaussian beam spot with a radius of approximately $10~\mathrm{\mu m}$. In contrast, the typical diameter of an e-beam spot is less than $100~\mathrm{nm}$. This significant dimensional mismatch necessitates a rigorous evaluation of the calibration validity. Consequently, Extended Data Fig.~\ref{Ex4} and the \textcolor{blue}{Supplementary Note 4} provide evidence that the size of the particle beam (the radius over which the force is applied) has a negligible impact on the optical pressure and the final calibration results. The detection optical path of the FINEST probe remains consistent with that shown in Fig.~\ref{Fig1}b. When the laser irradiates the top plate, photons continuously exert a vertically downward force on the FINEST probe. This phenomenon arises from the wave-particle duality of photons and the principle of momentum conservation. This applied force consequently compresses the helical structure (Fig.~\ref{Fig3}b).

Under the influence of optical pressure, the mechanical deformation of the helix is captured in real time as a spectral shift via optomechanical coupling. Fig.~\ref{Fig3}c through e displays the acquired spectral shifts under laser irradiation at powers of $1~\mathrm{mW}$, $10~\mathrm{mW}$, and $20~\mathrm{mW}$. It is evident that higher optical power induces a more pronounced spectral shift. Furthermore, the optical power exhibits a strict linear correlation with the wavelength shift ($\Delta\lambda$). A custom Python program was developed to capture the spectral resonance peaks and calculate the wavelength shift distance (see \textcolor{blue}{Methods \ref{8.4}}). This program converts the spectral information acquired by the optical power meter into $\Delta\lambda$ in real time.

Following the establishment of the relationship between laser power and spectral shift, the absolute magnitude of the laser-induced force must be determined. Optical radiation pressure is directly correlated with power. The conversion between laser power and thrust is governed by the radiation pressure equation, $F = (A + 2R) \frac{W}{c}$. Here, $A$ and $R$ represent the absorptance and reflectance of the metal film. The variable $W$ is the laser power, and $c$ is the speed of light. In this experiment, optical transmission through the top plate is negligible. This boundary condition renders $A + R = 1$. Consequently, the original equation simplifies to:
\begin{linenomath*}
\begin{equation}\label{eq:guangli}
F = (1+R) \frac{W}{c}.
\end{equation}
\end{linenomath*}
Therefore, an accurate measurement of the reflectance of the probe's Ag mirror top plate is critical for determining the laser radiation pressure. Reflectance is highly dependent on the thickness of the deposited Ag film (Fig.~\ref{Fig3}f). Experimentally deducing the film thickness from the natural frequency shift of a quartz crystal microbalance, and subsequently simulating the top plate reflectance, introduces inaccuracies. To address this, direct characterizations were performed using a spectrometer. Following the experiments, the top plate was sectioned (see the inset of Fig.~\ref{Fig3}f, Extended Data Fig.~\ref{Ex2}, and \textcolor{blue}{Supplementary Note 5}) and measured under an SEM. The results closely match the simulations.

This establishes the precise physical relationship between the laser power and the exerted optical pressure. The power of the single-frequency laser is continuously adjustable. To generate varying optical pressures, the laser power was systematically increased from $1~\mathrm{mW}$ to $22~\mathrm{mW}$ in increments of $1~\mathrm{mW}$. The corresponding resonance wavelength shifts $\Delta\lambda$ at each power level were recorded. These data are plotted in Fig.~\ref{Fig3}g. Subsequently, the radiation pressure equation was utilized to convert the incident laser power into absolute radiation pressure. This converted value corresponds to the vertical axis in Fig.~\ref{Fig3}h. Furthermore, the shift in the interference spectrum was translated into the change in the F-P cavity length $\Delta L$. This translation is governed by the following equation:

\begin{linenomath*}
\begin{equation}\label{eq:saunli}
\Delta L = \frac{L}{\lambda} \Delta \lambda.
\end{equation}
\end{linenomath*}

The resulting cavity length variations are plotted in Fig.~\ref{Fig3}h. The physical significance of the linear fitting slope in this plot corresponds directly to the stiffness of the FINEST probe. Based on this slope, the spring constant ($k$) of the probe was calculated to be $0.00078~\mathrm{N~m^{-1}}$. Utilizing this calibrated stiffness, the ultimate sensing performance of the FINEST platform was quantitatively evaluated. The system exhibits an absolute mechanical sensitivity of $13.35~\mathrm{pm~pN^{-1}}$. Furthermore, bounded by a free spectral range of approximately $8.13$~nm, the platform affords a broad theoretical dynamic range of $609$~pN. Coupled with an ultra-low static baseline fluctuation (standard deviation of $0.036$~pm), these metrics confirm the stringent resolution capability required for exploring femtonewton-scale electron momentum-transfer mechanisms. Additionally, the temperature cross-sensitivity of the FINEST probe was investigated (see Extended Data Fig.~\ref{Ex3} and \textcolor{blue}{Supplementary Note 6}). The measured sensitivity is 19.1~pm/$^\circ$C. This remarkably low value demonstrates the excellent thermal robustness of the probe.

\section{Benchmarking force metrology in multiphysics environments}

To elucidate the physical scope of this work, a comprehensive comparison was conducted. This maps typical natural forces against existing sensing technologies (Fig.~\ref{Fig3}i). Following calibration, the developed FINEST platform pushes the limit of fiber-integrated all-optical force sensors down to the fN scale. This securely covers the measurement range of e-beam pressure.

However, achieving this measurement limit inside an electron microscope is not merely a competition of sensitivity. It requires a comprehensive assessment against extreme multiphysics environments. As summarized in Table 1 and Fig.~\ref{Fig3}i, the overall performance of state-of-the-art mechanical sensing technologies was systematically evaluated within a low-vacuum electron microscope environment. Although optical tweezers\cite{liu2016subfemtonewton} and atomic force microscopy (AFM)\cite{melcher2014self} possess fN/pN-level detection capabilities under standard conditions, they fail to operate or face spatial integration challenges in extremely low-pressure environments and confined in-situ chambers. Conversely, conventional MEMS sensors based on capacitance, piezoresistance, or piezoelectricity\cite{han2023advances} are highly susceptible to strong electromagnetic interference (EMI) and severe surface charge accumulation under direct e-beam bombardment. These effects lead to complete readout signal distortion.

In contrast, the FINEST platform exhibits distinct advantages under all these stringent physical constraints. First, the purely optical F-P interferometric readout mechanism is inherently immune to any form of EMI. Second, the custom Ag film coating on the probe surface works in synergy with the active temperature control strategy. This combination resolves the thermal drift issue of the polymer structure. Furthermore, it establishes an excellent active charge dissipation channel, effectively eliminating electrostatic repulsion artifacts caused by electron accumulation. Combined with the high in-situ integrability afforded by its compact fiber facet architecture, the FINEST stands as a unique mechanical sensing solution. It is currently the only platform capable of simultaneously offering vacuum compatibility, robust anti-interference capability, and an fN-level ultimate detection threshold within complex multiphysics coupled environments.

\begin{table*}[htbp]
    \centering
    \caption{Comparative evaluation of state-of-the-art force sensing technologies in vacuum electron microscopy environments.}
    \label{tab:comparative_evaluation}
    
    \resizebox{\textwidth}{!}{ 
    \begin{tabular}{lccccccc}
        \toprule
        \makecell[l]{Sensing Technology \textbackslash \\ Environmental Constraint} & 
        \makecell{Capacitive\\MEMS} & 
        \makecell{Piezoresistive\\MEMS} & 
        \makecell{Piezoelectric\\MEMS} & 
        AFM & 
        OT & 
        \makecell{Conventional\\Fiber Probe} & 
        This work \\
        \midrule
        Vacuum/low-pressure       & \checkmark & \checkmark & \checkmark & \checkmark  & $\times$   & \checkmark & \checkmark \\
        EMI immunity              & $\times$   & $\times$   & $\times$   & $\triangle$ & \checkmark & \checkmark & \checkmark \\
        Active charge dissipation & $\times$   & $\times$   & $\times$   & $\triangle$ & $\times$   & $\times$   & \checkmark \\
        Thermal drift immunity    & $\triangle$& $\times$   & \checkmark & \checkmark  & $\times$   & \checkmark & \checkmark \\
        Detection limit           & $\times$   & $\times$   & $\triangle$& $\triangle$ & \checkmark & $\triangle$& \checkmark \\
        In-situ integrability     & $\triangle$& $\triangle$& $\triangle$& $\triangle$ & $\times$   & \checkmark & \checkmark \\
        Overall suitability       & $\times$   & $\times$   & $\times$   & $\times$    & $\times$   & $\times$   & \checkmark \\
        \bottomrule
    \end{tabular}
    } 
    
    \vspace{1.5ex}
    
    \begin{minipage}{\textwidth}
        \small Legend: \checkmark{} Fully satisfies the criterion; $\times$ Fails to satisfy the criterion; $\triangle$ Conditionally or partially satisfies the criterion (highly dependent on specific structural implementation or necessitates a functional trade-off).
    \end{minipage}
\end{table*}

\section{\textit{In situ} mechanics of electron momentum}

\begin{figure}[htbp]
\centering
\includegraphics[width=1\textwidth]{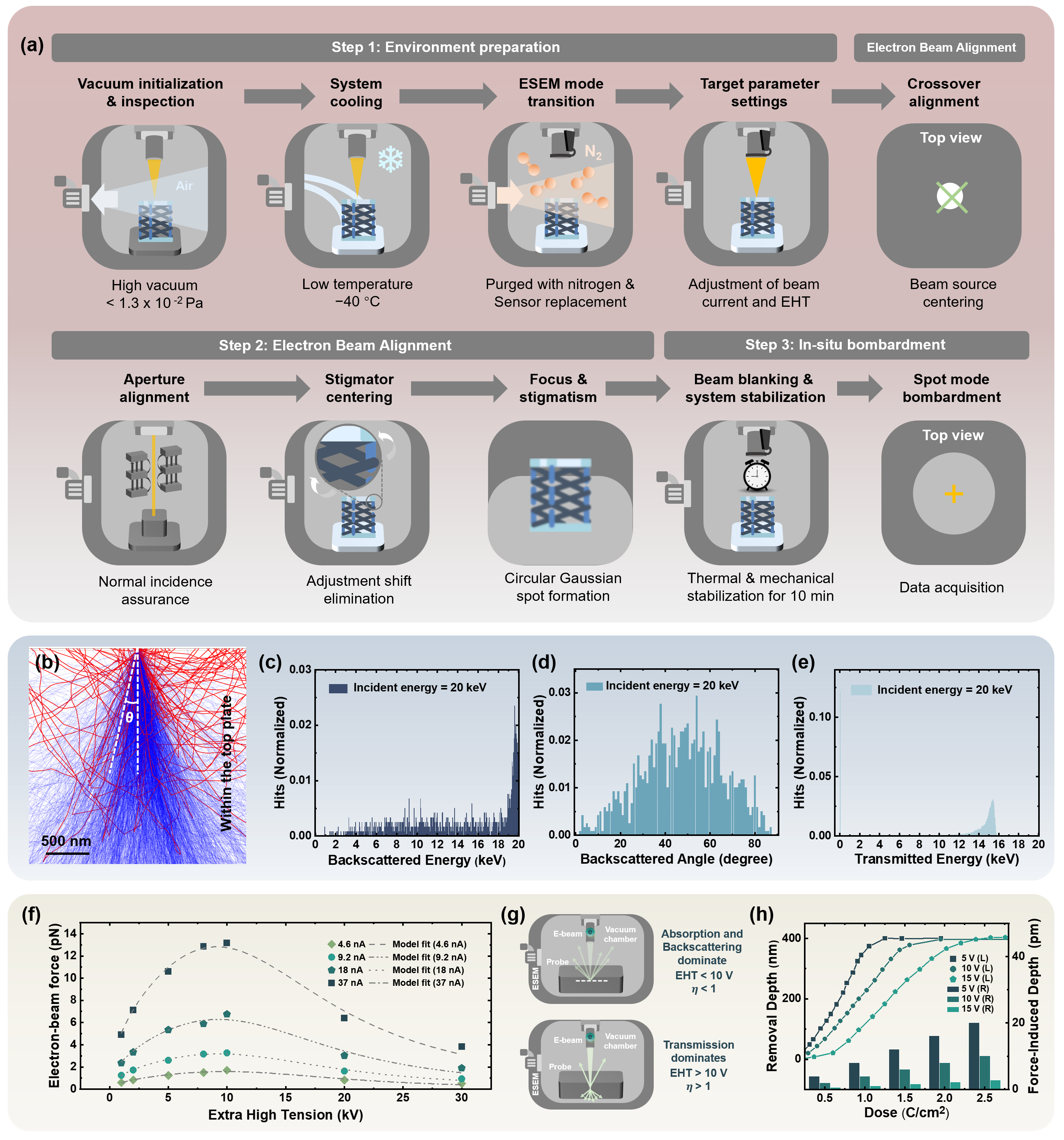}
\caption{\textbf{Direct measurement of electron-beam momentum-transfer force.} \textbf{(a)} Schematic workflow for \textit{in situ} electron-beam bombardment force measurements and force readout. \textbf{(b)} Monte Carlo trajectory simulation of 20~keV incident electrons in the top plate of the probe, showing the propagation of incident and transmitted electrons (both in blue) and backscattered electrons (red). Here, $\theta$ represents the exit angle of the transmitted electrons. Scale bar, 500~nm. \textbf{(c)} Histogram of the backscattered-electron energy distribution at 20~keV. \textbf{(d)} Histogram of the backscattered-electron angular distribution at 20~keV. \textbf{(e)} Histogram of the transmitted-electron energy distribution at 20~keV. \textbf{(f)} Electron-beam force as a function of EHT under different beam-current conditions. Symbols denote measured forces, and dashed curves denote model fits. The force exhibits a non-monotonic dependence on EHT, reaching a maximum at intermediate voltages. \textbf{(g)} Schematic of the interaction mechanism between the e-beam and the probe within the ESEM chamber. At different EHT values, electron absorption and backscattering processes, and electron transmission behavior dominate, respectively. \textbf{(h)} Ice removal under electron-beam irradiation and analysis of its mechanical contribution. Ice-removal depth as a function of electron dose is shown by lines (left axis) at different EHT values, and force-induced depth converted from the measured electron-beam force is shown by bars (right axis). The removal depth increases monotonically with dose, whereas the force-induced depth remains far smaller than the measured removal depth, indicating that pure mechanical momentum transfer is insufficient to account for electron-beam-induced ice removal.}
\label{Fig4}
\end{figure}

Following the optical calibration, the FINEST probe was utilized to quantitatively measure the electron beam-induced force within an environmental scanning electron microscope (ESEM). However, extracting the authentic mechanical signal demands more than baseline force sensitivity. Resonance shifts can also originate from thermal drift, beam misalignment, charging effects, or environmental fluctuations within the chamber. To address this, a staged \textit{in situ} bombardment protocol was established (Fig.~\ref{Fig4}a). This procedure strictly controls the environmental, optical, and e-beam conditions prior to any force readout.

First, the probe is inspected under high vacuum. It is then cooled to the target temperature before the system is switched to the ESEM working mode\cite{danilatos1991review, danilatos1988foundations}. Subsequently, the chamber pressure is adjusted by introducing nitrogen gas. The e-beam current and acceleration voltage are explicitly set before each measurement. The measured axial response must correspond strictly to well-defined beam loading conditions. To ensure this, the electron column sequentially undergoes crossover centering, aperture alignment, astigmatism correction, and focus optimization. These steps produce a centered, nearly circular Gaussian beam spot. This optimized spot is then directed onto the top plate of the helical probe. Prior to data acquisition, the e-beam is blanked. The FINEST system is then allowed to stabilize for $10~\mathrm{min}$. Finally, the e-beam irradiates the center of the top plate in spot mode. Simultaneously, the Fabry-Pérot resonance signal is continuously recorded in real time.

This workflow directly links the preset e-beam conditions with the calibrated mechanical readout. It effectively minimizes the contribution of non-mechanical factors to the measured resonance shift. Consequently, it provides a robust experimental foundation. This foundation is essential for extracting the force scaling relationships with beam current and acceleration voltage, and for comparing these experimental trends against electron transport simulations.

To select appropriate EHT and beam current parameters during the experiments and provide a comprehensive theoretical foundation for the interaction process, a momentum-transfer model relying on Monte Carlo statistical expectations was developed based on the conservation of boundary momentum flux. The detailed derivation of this model is provided in the \textcolor{blue}{Supplementary Note 7}. This yields the governing equation for the macroscopic force $F$, which introduces a dimensionless correction factor $\eta$:
\begin{linenomath*}
\begin{equation}\label{eq:1}
F = \eta \frac{I}{e} P_0 = \frac{I}{e} P_0 \left[ 1 + f_B \left\langle \sqrt{\frac{E_B}{E_0}} \cos\theta_B \right\rangle_B - f_T \left\langle \sqrt{\frac{E_T}{E_0}} \cos\theta_T \right\rangle_T \right].
\end{equation}
\end{linenomath*}
Here, $\eta$ represents the momentum transfer efficiency governed by the microscopic scattering events. The variable $I$ is the beam current, and $e$ represents the elementary charge. The variable $P_0$ denotes the initial normal momentum of the incident electrons. The dimensionless terms $f_B$ and $f_T$ correspond to the fractions of backscattered and transmitted electrons, respectively. The variables $E_0$, $E_B$, and $E_T$ represent the incident energy, the emission energy of backscattered electrons, and the emission energy of transmitted electrons. Furthermore, $\theta_B$ and $\theta_T$ denote the respective emission angles of these two electron populations relative to the surface normal.

The statistical results of these variables are presented in Fig.~\ref{Fig4}b-e and Extended Data Fig.~\ref{Ex7} via Monte Carlo simulations. The \textcolor{blue}{Supplementary Note 8} provides a detailed analysis of the simulation results in conjunction with the physical model. Furthermore, the \textcolor{blue}{Methods \ref{8.5}} section elaborates on the specific simulation procedures. At an incident energy of $20~\mathrm{keV}$, both the backscattered and transmitted energy spectra exhibit a sharp peak near the initial incident energy. Simultaneously, the backscattered angle distribution displays a broad angular spread. By integrating these statistical data with Equation \eqref{eq:1}, the magnitude of the electron pressure under specific parameters can be approximately estimated.

Guided by the analysis of the aforementioned model and the calibrated performance parameters of the FINEST probe, the beam current was systematically adjusted between $4.6~\mathrm{nA}$ and $37~\mathrm{nA}$. Concurrently, the EHT was tuned between $1~\mathrm{kV}$ and $30~\mathrm{kV}$. This configuration generated dozens of two-parameter combinations. For each combination, repeated experiments were conducted to acquire multiple spectral resonance shift results. These results were subsequently averaged to obtain the final e-beam pressure data. Notably, the low-temperature and vacuum environments significantly enhanced both the stability and quality of the interference spectrum (see Extended Data Fig.~\ref{Ex3}c). 

Following these experimental procedures, a scatter plot of the e-beam pressure corresponding to different EHT and beam currents is presented in Fig.~\ref{Fig4}f. A phenomenological scaling model for electron momentum transfer was constructed to fit the experimental data. The resulting fit is indicated by the gray dashed lines. This analytical expression is formulated as:
\begin{linenomath*}
\begin{equation}
F = A \sqrt{E} \left[ \frac{1}{1 + (E/E_0)^n} \right].
\end{equation}
\end{linenomath*}
This model successfully fits all currently available experimental results. The specific construction process and physical implications are detailed in the \textcolor{blue}{Supplementary Note 9}. The magnitudes of these experimentally measured forces span from $505~\mathrm{fN}$ to $13~\mathrm{pN}$. Analyzing these data yields two significant conclusions based on quantitative evaluation.
 
The first conclusion aligns with physical intuition. As the e-beam current increases, the e-beam pressure increases accordingly. Furthermore, experimental results demonstrate a strict linear proportionality between the e-beam pressure and the beam current under a constant EHT (Fig.~\ref{Fig4}f). This is consistent with the theoretical derivation of Equation \eqref{eq:1}. This phenomenon was also noted in the simulations, showing a linear relationship between $I$ and $F$. From a microscopic perspective, an increase in the beam current corresponds directly to a proportional increase in the incident electron flux per unit time. Under a constant EHT, the initial momentum carried by each individual electron remains unchanged. Simultaneously, the statistical probabilities determining the electron transport pathways, such as backscattering and transmission, remain unchanged. The macroscopic force manifests as a linear temporal accumulation of discrete single-electron momentum transfer events. This observation experimentally verifies the statistical validity of momentum conservation during massive electron-matter interactions.

The second conclusion reflects the nonlinear dependence of the e-beam pressure on the EHT. When the e-beam current is constant and only the EHT is varied, the e-beam pressure exhibits a trend of initial increase followed by a decrease. This trend produces a maximum force value near $10~\mathrm{kV}$. This phenomenon was observed across different beam currents. Combining the simulations and the physical model, this is attributed to the competing mechanisms of the electron absorption, backscattering, and transmission processes (see Fig.~\ref{Fig4}g). The nonlinear evolution of the correction factor $\eta$ was introduced into the model to quantitatively characterize this process (see Equation \eqref{eq:1} and \textcolor{blue}{Supplementary Notes 7 and 8}). In the low-energy regime, almost no e-beam can penetrate the probe. The electron absorption and backscattering processes dominate, resulting in $\eta > 1$. The probe detects an additional recoil force exceeding the pure absorption limit. In the high-energy regime, a considerable proportion of electrons penetrates the probe. The electron transmission processes dominate, resulting in $\eta < 1$. Momentum is lost. Consequently, the e-beam pressure detected by the probe decreases rapidly.

Although similar non-monotonic trends have been qualitatively observed in traditional energy deposition simulations\cite{assa2018monte}, the exact mechanical values have remained unknown. This knowledge gap persists because, while idealized theoretical formalisms for electron-induced momentum transfer\cite{garcia2004momentum} exist, the lack of experimental validation\cite{susi2019quantifying} has hindered the direct quantitative extraction of the momentum vector flux. Through this direct measurement, we quantified previously undetermined physical values. This work supplements a long-missing dynamic dimension for fields such as e-beam imaging, precision manufacturing, and space propulsion.

To demonstrate the practical implications of this newly quantified dimension, a paradigmatic complex system was selected for in-situ investigation. The interaction between e-beams and amorphous ice is a fundamental physical process with profound interdisciplinary significance. In astrophysics, it simulates the cosmic weathering of icy planetary bodies\cite{mifsud2022comparative}. In structural biology, amorphous ice serves as the standard embedding medium for cryo-electron microscopy\cite{adrian1984cryo}. In nanotechnology, it is utilized as a promising resist for in vivo and three-dimensional nanofabrication\cite{hong2018three}. Despite its crucial importance, the material removal mechanism during the electron-ice interaction remains highly complex. It involves an entangled combination of thermal sublimation, chemical radiolysis, and mechanical momentum transfer\cite{abellan2023radiolysis, xu2020electron}. Although thermal and chemical pathways have been extensively investigated, the exact contribution of pure mechanical momentum transfer remains unknown. The existence of this gap is primarily due to the prior lack of femtonewton-scale in-situ measurement tools.

Within a custom scanning electron microscope chamber cooled to $130~\mathrm{K}$ by liquid nitrogen, high-purity water vapor was utilized as a gas-phase precursor. Dense and thickness-controllable amorphous solid water resist films were deposited and grown in situ on targeted silicon substrates. Subsequently, the amorphous ice was patterned and etched using an e-beam\cite{wu2021lithographic} (see \textcolor{blue}{Methods \ref{8.6}}). The evolution of etching depth versus electron dose was recorded (line plots in Fig.~\ref{Fig4}h). Extended Data Fig.~\ref{Ex8} documents the etching patterns under varying voltages and beam currents dictated by different apertures. The etching thickness was acquired via SEM imaging after tilting the displacement stage by $30^\circ$ (Extended Data Fig.~\ref{Ex9}a,b). The theoretical relationships among parameters including the e-beam EHT, beam current, and dose are illustrated in Extended Data Fig.~\ref{Ex9}c. As derived from Fig.~\ref{Fig4}h, the observations reveal a distinct energy dependence in etching efficiency. Specifically, more significant ice layer removal occurs at a lower acceleration voltage of $5~\mathrm{kV}$.

Based on the momentum-flux model for electron-beam force and the preceding experimental data, this study provides the first quantitative calculation of the theoretical contribution from pure mechanical momentum transfer to ice removal (for the calculation procedure, see \textcolor{blue}{Supplementary Note 10}). These calculated values are displayed as bar charts in Fig.~\ref{Fig4}h. A quantitative comparison reveals that the experimentally measured total etching depth far exceeds the upper limit of the purely mechanically induced depth. The difference spans more than four orders of magnitude. Furthermore, the scaling trend of the mechanical contribution with acceleration voltage is inconsistent with the energy dependence of the actual etching depth. These results indicate that mechanical removal driven by momentum transfer is not the dominant mechanism during electron beam-induced ice etching. Its physical effect is fundamentally negligible. This application demonstrates the critical importance of directly measuring the e-beam momentum transfer force. Such measurements are essential for distinguishing mechanical effects from other energy deposition processes within complex electron-matter interactions.

\section{Discussion and outlook}
To our knowledge, this work provides the first experimental absolute calibration of electron-beam momentum transfer utilizing FINEST technology, establishing a statistical correlation between microscopic discrete single-electron scattering and macroscopic continuous mechanical responses. The non-monotonic trend of the mechanical effects with varying acceleration voltages, as revealed in the experiments, provides quantitative physical constraints for understanding energy dissipation and momentum allocation during electron-matter interactions. Building upon this foundation, electron-beam etching experiments on amorphous water ice confirm that the absolute contribution of purely mechanical momentum transfer to material removal is exceedingly minimal. These findings potentially offer a reliable quantitative reference, from the perspective of non-thermal phase transition dynamics, for evaluating radiation damage in structural biology cryo-media and understanding the weathering evolution of interstellar ice bodies in astrophysics.

This technology exhibits potential for further advancement regarding material scalability, environmental compatibility, and detection limits. A distinct advantage of the FINEST platform lies in the customizability of the probe target material. Current experiments primarily rely on silver film coatings. Future configurations could integrate target materials of varying thicknesses, shapes, lattice structures, and atomic numbers onto the probe top plate, such as other metals, specific semiconductor substrates, or 2D materials. This high degree of scalability allows the system to transition into a versatile in situ platform for systematically studying momentum transfer variations during interactions between e-beams and diverse substrates, thereby providing a novel nanoscale characterization method for e-beam effects.

The environmental compatibility of the detector is anticipated to be further enhanced through advancements in two-photon 3D printing technology. Constrained by the electrical and thermal conductivities of polymer materials, the probe currently relies on surface metallization, ESEM modes, and active cooling to maintain reliable performance. Recent studies have successfully developed printing materials incorporated with metallic particles\cite{han2022three}. Should the fabrication of fully metallic micro-helical springs be realized in the future, these probes hold the potential to operate stably within standard high-vacuum electron microscopy environments. This capability would better align with practical physical application scenarios, including advanced semiconductor manufacturing and precision measurement.

Furthermore, the mechanical sensitivity of the FINEST technology holds potential for further enhancement. Currently, the geometric dimensions of the probe are constrained by the effective fabrication field of view of the objective lens in the commercial two-photon lithography system. In the future, utilizing a low-magnification objective lens with a larger field of view\cite{obata2013high, gu20253d} could expand the radius of the helical spring while maintaining fabrication quality. Alternatively, enhancing the ultimate spatial resolution of the printing process could enable the fabrication of helical beams with sub-micron cross-sectional dimensions. Both approaches would effectively reduce the spring constant of the probe, thereby significantly improving the system's responsiveness to minute forces.

The fabrication strategy employed for the FINEST probe is inherently generalizable. To overcome the challenges associated with surface tension, we implemented a six-step process optimization, featuring architecture-inspired scaffold support structures, and conducted rigorous mechanical modeling. Analogous fabrication challenges are prevalent in the manufacturing of microelectromechanical systems (MEMS) and nanoelectromechanical systems (NEMS)\cite{mastrangelo2002mechanical, namatsu1995dimensional}, particularly during the wet release of sacrificial layers in silicon-based processing and the liquid-phase cleaning and drying of high-aspect-ratio microstructures. This process optimization strategy is anticipated to be equally applicable to the fabrication of such complex structures. Consequently, this could facilitate the practical application of extreme-geometry micro- and nanoarchitectures in micro- and nanorobotics, biomedical interfaces, ultra-precision metrology, and broader fields.

As a general methodology for fundamental mechanical measurements, this platform exhibits significant potential for synergy with other advanced technologies. By incorporating sideband cooling techniques\cite{teufel2011sideband} within the optomechanical architecture, the probe could potentially be cooled to its quantum ground state. This would provide an experimental avenue for observing macroscopic decoherence mechanisms induced by microscopic particle bombardment. Furthermore, integration with heavy-particle emission sources, such as focused ion beams\cite{li2021recent}, enables the measurement of absolute momentum transfer accompanying ion implantation. This capability could offer mechanical guidance for evaluating lattice damage and precisely delineating thresholds for non-destructive processing.

The quantification of electron-beam-induced forces achieved in this work opens up opportunities to expand the applications of e-beams themselves.  Beyond imaging and fabrication, the e-beam is envisioned as a non-contact nanomechanical probe for precise calibration. This conceptualization of the e-beam as an in-situ force source provides a new perspective for the evolution of electron microscopy instrumentation.  By enabling non-contact mechanical manipulation, this approach effectively circumvents the contact wear and surface-force interferences inherently introduced by traditional mechanical probes during the characterization of fragile specimens.  Concurrently, leveraging the targeted bombardment capabilities of nanoscale beam spots facilitates the development of localized mechanical mapping techniques with high spatial resolution.  Furthermore, benefiting from the physical advantage of lacking mass-inertial hysteresis, the e-beam, when combined with high-frequency beam modulation, holds the potential to apply alternating loads in the $\mathrm{MHz}$ range to target samples\cite{huber2022tailoring}, thereby serving as a viable high-frequency excitation source for in-situ dynamic mechanical analysis.  This emerging electron-beam-based characterization paradigm is anticipated to advance fundamental research and pioneering applications across condensed-matter physics, structural biology, and micro- and nanomanufacturing.

In summary, the FINEST technology presented herein resolves the long-standing experimental challenge of quantitatively measuring electron-beam momentum and induced forces. It successfully achieves the capture of femtonewton ($\mathrm{fN}$)-scale forces under strong electronic interference within an in-situ electron-beam environment. Furthermore, building upon these experimental measurements, this study offers theoretical refinements and elucidations from the perspectives of beam current and energy dependencies. This work thereby integrates a long-missing mechanical dimension into the physical picture of electron-beam processes. With continuous advancements in measurement resolution and system integration, we predict that the FINEST platform and its derivative nanomechanical sensing technologies will accelerate the thorough exploration of momentum effects inherent in the wave-particle duality of electrons, thereby facilitating practical applications across next-generation semiconductor manufacturing, quantum-material development, deep-space exploration, structural biology cryo-microscopy, and broader interdisciplinary fields.

\section*{Online content}

Any methods, additional references, Nature Portfolio reporting summaries, source data, extended data, supplementary information, acknowledgements, peer review information; details of author contributions and competing interests; and statements of data and code availability are available at \url{URL}.

\renewcommand{\refname}{References}
\begin{flushleft}
\bibliography{\jobname}

\begin{thebibliography}{10}
\expandafter\ifx\csname url\endcsname\relax
  \def\url#1{\burl{#1}}\fi
\expandafter\ifx\csname urlprefix\endcsname\relax\def\urlprefix{URL }\fi
\providecommand{\bibinfo}[2]{#2}
\providecommand{\eprint}[2][]{\url{#2}}
\providecommand{\doi}[1]{\url{https://doi.org/#1}}
\bibcommenthead

\bibitem{obayashi1984space}
\bibinfo{author}{Obayashi, T.} \emph{et~al.}
\newblock \bibinfo{title}{Space experiments with particle accelerators}.
\newblock \emph{\bibinfo{journal}{Science}} \textbf{\bibinfo{volume}{225}}, \bibinfo{pages}{195--196} (\bibinfo{year}{1984}) .

\bibitem{deacon1977first}
\bibinfo{author}{Deacon, D.~A.} \emph{et~al.}
\newblock \bibinfo{title}{First operation of a free-electron laser}.
\newblock \emph{\bibinfo{journal}{Phys. Rev. Lett.}} \textbf{\bibinfo{volume}{38}}, \bibinfo{pages}{892} (\bibinfo{year}{1977}) .

\bibitem{litos2014high}
\bibinfo{author}{Litos, M.} \emph{et~al.}
\newblock \bibinfo{title}{High-efficiency acceleration of an electron beam in a plasma wakefield accelerator}.
\newblock \emph{\bibinfo{journal}{Nature}} \textbf{\bibinfo{volume}{515}}, \bibinfo{pages}{92--95} (\bibinfo{year}{2014}) .

\bibitem{wkeglowski2016electron}
\bibinfo{author}{W{\k{e}}glowski, M.~S.}, \bibinfo{author}{B{\l}acha, S.} \& \bibinfo{author}{Phillips, A.}
\newblock \bibinfo{title}{Electron beam welding--techniques and trends--review}.
\newblock \emph{\bibinfo{journal}{Vacuum}} \textbf{\bibinfo{volume}{130}}, \bibinfo{pages}{72--92} (\bibinfo{year}{2016}) .

\bibitem{von1938elektronen}
\bibinfo{author}{Von~Ardenne, M.}
\newblock \bibinfo{title}{Das elektronen-rastermikroskop: Theoretische grundlagen}.
\newblock \emph{\bibinfo{journal}{Z. Phys.}} \textbf{\bibinfo{volume}{109}}, \bibinfo{pages}{553--572} (\bibinfo{year}{1938}) .

\bibitem{huth2012focused}
\bibinfo{author}{Huth, M.} \emph{et~al.}
\newblock \bibinfo{title}{Focused electron beam induced deposition: A perspective}.
\newblock \emph{\bibinfo{journal}{Beilstein J. Nanotechnol.}} \textbf{\bibinfo{volume}{3}}, \bibinfo{pages}{597--619} (\bibinfo{year}{2012}) .

\bibitem{broers1976electron}
\bibinfo{author}{Broers, A.}, \bibinfo{author}{Molzen, W.}, \bibinfo{author}{Cuomo, J.} \& \bibinfo{author}{Wittels, N.}
\newblock \bibinfo{title}{Electron-beam fabrication of 80-{\aa} metal structures}.
\newblock \emph{\bibinfo{journal}{Appl. Phys. Lett.}} \textbf{\bibinfo{volume}{29}}, \bibinfo{pages}{596--598} (\bibinfo{year}{1976}) .

\bibitem{drouin2007casino}
\bibinfo{author}{Drouin, D.} \emph{et~al.}
\newblock \bibinfo{title}{Casino v2.42—a fast and easy-to-use modeling tool for scanning electron microscopy and microanalysis users}.
\newblock \emph{\bibinfo{journal}{Scanning}} \textbf{\bibinfo{volume}{29}}, \bibinfo{pages}{92--101} (\bibinfo{year}{2007}) .

\bibitem{susi2019quantifying}
\bibinfo{author}{Susi, T.}, \bibinfo{author}{Meyer, J.~C.} \& \bibinfo{author}{Kotakoski, J.}
\newblock \bibinfo{title}{Quantifying transmission electron microscopy irradiation effects using two-dimensional materials}.
\newblock \emph{\bibinfo{journal}{Nat. Rev. Phys.}} \textbf{\bibinfo{volume}{1}}, \bibinfo{pages}{397--405} (\bibinfo{year}{2019}) .

\bibitem{pairis2019shot}
\bibinfo{author}{Pairis, S.} \emph{et~al.}
\newblock \bibinfo{title}{Shot-noise-limited nanomechanical detection and radiation pressure backaction from an electron beam}.
\newblock \emph{\bibinfo{journal}{Phys. Rev. Lett.}} \textbf{\bibinfo{volume}{122}}, \bibinfo{pages}{083603} (\bibinfo{year}{2019}) .

\bibitem{fiorillo2018theory}
\bibinfo{author}{Fiorillo, A.~S.}, \bibinfo{author}{Critello, C.~D.} \& \bibinfo{author}{Pullano, S.~A.}
\newblock \bibinfo{title}{Theory, technology and applications of piezoresistive sensors: A review}.
\newblock \emph{\bibinfo{journal}{Sens. Actuators A Phys.}} \textbf{\bibinfo{volume}{281}}, \bibinfo{pages}{156--175} (\bibinfo{year}{2018}) .

\bibitem{mishra2021recent}
\bibinfo{author}{Mishra, R.~B.}, \bibinfo{author}{El-Atab, N.}, \bibinfo{author}{Hussain, A.~M.} \& \bibinfo{author}{Hussain, M.~M.}
\newblock \bibinfo{title}{Recent progress on flexible capacitive pressure sensors: From design and materials to applications}.
\newblock \emph{\bibinfo{journal}{Adv. Mater. Technol.}} \textbf{\bibinfo{volume}{6}}, \bibinfo{pages}{2001023} (\bibinfo{year}{2021}) .

\bibitem{vashist2007review}
\bibinfo{author}{Vashist, S.~K.}
\newblock \bibinfo{title}{A review of microcantilevers for sensing applications}.
\newblock \emph{\bibinfo{journal}{J. Nanotechnol.}} \textbf{\bibinfo{volume}{3}}, \bibinfo{pages}{36} (\bibinfo{year}{2007}) .

\bibitem{liu2020integrated}
\bibinfo{author}{Liu, T.} \emph{et~al.}
\newblock \bibinfo{title}{Integrated nano-optomechanical displacement sensor with ultrawide optical bandwidth}.
\newblock \emph{\bibinfo{journal}{Nat. Commun.}} \textbf{\bibinfo{volume}{11}}, \bibinfo{pages}{2407} (\bibinfo{year}{2020}) .

\bibitem{binnig1986atomic}
\bibinfo{author}{Binnig, G.}, \bibinfo{author}{Quate, C.~F.} \& \bibinfo{author}{Gerber, C.}
\newblock \bibinfo{title}{Atomic force microscope}.
\newblock \emph{\bibinfo{journal}{Phys. Rev. Lett.}} \textbf{\bibinfo{volume}{56}}, \bibinfo{pages}{930} (\bibinfo{year}{1986}) .

\bibitem{jones2015optical}
\bibinfo{author}{Jones, P.}, \bibinfo{author}{Marag{\'o}, O.} \& \bibinfo{author}{Volpe, G.}
\newblock \emph{\bibinfo{title}{Optical Tweezers: Principles and Applications}}  (\bibinfo{publisher}{Cambridge University Press}, \bibinfo{address}{Cambridge}, \bibinfo{year}{2015}).

\bibitem{li2020ultrathin}
\bibinfo{author}{Li, J.} \emph{et~al.}
\newblock \bibinfo{title}{Ultrathin monolithic 3d printed optical coherence tomography endoscopy for preclinical and clinical use}.
\newblock \emph{\bibinfo{journal}{Light Sci. Appl.}} \textbf{\bibinfo{volume}{9}}, \bibinfo{pages}{124} (\bibinfo{year}{2020}) .

\bibitem{zou2021fiber}
\bibinfo{author}{Zou, M.} \emph{et~al.}
\newblock \bibinfo{title}{Fiber-tip polymer clamped-beam probe for high-sensitivity nanoforce measurements}.
\newblock \emph{\bibinfo{journal}{Light Sci. Appl.}} \textbf{\bibinfo{volume}{10}}, \bibinfo{pages}{171} (\bibinfo{year}{2021}) .

\bibitem{li2025fish}
\bibinfo{author}{Li, L.} \emph{et~al.}
\newblock \bibinfo{title}{From fish to fiber: 3d-nanoprinted optical neuromast for multi-integrated underwater detection}.
\newblock \emph{\bibinfo{journal}{Nat. Commun.}} \textbf{\bibinfo{volume}{16}}, \bibinfo{pages}{7390} (\bibinfo{year}{2025}) .

\bibitem{shang2024fiber}
\bibinfo{author}{Shang, X.} \emph{et~al.}
\newblock \bibinfo{title}{Fiber-integrated force sensor using 3d printed spring-composed fabry-perot cavities with a high precision down to tens of piconewton}.
\newblock \emph{\bibinfo{journal}{Adv. Mater.}} \textbf{\bibinfo{volume}{36}}, \bibinfo{pages}{2305121} (\bibinfo{year}{2024}) .

\bibitem{shang2022customizable}
\bibinfo{author}{Shang, X.} \emph{et~al.}
\newblock \bibinfo{title}{Customizable and highly sensitive 3d micro-springs produced by two-photon polymerizations with improved post-treatment processes}.
\newblock \emph{\bibinfo{journal}{Appl. Phys. Lett.}} \textbf{\bibinfo{volume}{120}} (\bibinfo{year}{2022}) .

\bibitem{delrio2005role}
\bibinfo{author}{DelRio, F.~W.} \emph{et~al.}
\newblock \bibinfo{title}{The role of van der waals forces in adhesion of micromachined surfaces}.
\newblock \emph{\bibinfo{journal}{Nat. Mater.}} \textbf{\bibinfo{volume}{4}}, \bibinfo{pages}{629--634} (\bibinfo{year}{2005}) .

\bibitem{hunt1981review}
\bibinfo{author}{Hunt, B.} \& \bibinfo{author}{Fattal, S.~G.}
\newblock \emph{\bibinfo{title}{Review of technical information on scaffolds}}  (\bibinfo{publisher}{National Bureau of Standards}, \bibinfo{address}{Washington, D.C.}, \bibinfo{year}{1981}).

\bibitem{martinez2017inertial}
\bibinfo{author}{Mart{\'\i}nez-Mart{\'\i}n, D.} \emph{et~al.}
\newblock \bibinfo{title}{Inertial picobalance reveals fast mass fluctuations in mammalian cells}.
\newblock \emph{\bibinfo{journal}{Nature}} \textbf{\bibinfo{volume}{550}}, \bibinfo{pages}{500--505} (\bibinfo{year}{2017}) .

\bibitem{yu2024optical}
\bibinfo{author}{Yu, W.} \emph{et~al.}
\newblock \bibinfo{title}{Optical nanofiber-enabled self-detection picobalance}.
\newblock \emph{\bibinfo{journal}{ACS Photonics}} \textbf{\bibinfo{volume}{11}}, \bibinfo{pages}{2316--2323} (\bibinfo{year}{2024}) .

\bibitem{zhu2025gold}
\bibinfo{author}{Zhu, J.} \emph{et~al.}
\newblock \bibinfo{title}{Gold flake-enabled miniature capacitive picobalances}.
\newblock \emph{\bibinfo{journal}{Small Methods}} \textbf{\bibinfo{volume}{9}}, \bibinfo{pages}{2401640} (\bibinfo{year}{2025}) .

\bibitem{liu2016subfemtonewton}
\bibinfo{author}{Liu, L.}, \bibinfo{author}{Kheifets, S.}, \bibinfo{author}{Ginis, V.} \& \bibinfo{author}{Capasso, F.}
\newblock \bibinfo{title}{Subfemtonewton force spectroscopy at the thermal limit in liquids}.
\newblock \emph{\bibinfo{journal}{Phys. Rev. Lett.}} \textbf{\bibinfo{volume}{116}}, \bibinfo{pages}{228001} (\bibinfo{year}{2016}) .

\bibitem{melcher2014self}
\bibinfo{author}{Melcher, J.}, \bibinfo{author}{Stirling, J.}, \bibinfo{author}{Cervantes, F.~G.}, \bibinfo{author}{Pratt, J.~R.} \& \bibinfo{author}{Shaw, G.~A.}
\newblock \bibinfo{title}{A self-calibrating optomechanical force sensor with femtonewton resolution}.
\newblock \emph{\bibinfo{journal}{Appl. Phys. Lett.}} \textbf{\bibinfo{volume}{105}} (\bibinfo{year}{2014}) .

\bibitem{han2023advances}
\bibinfo{author}{Han, X.} \emph{et~al.}
\newblock \bibinfo{title}{Advances in high-performance mems pressure sensors: design, fabrication, and packaging}.
\newblock \emph{\bibinfo{journal}{Microsyst. Nanoeng.}} \textbf{\bibinfo{volume}{9}}, \bibinfo{pages}{156} (\bibinfo{year}{2023}) .

\bibitem{danilatos1991review}
\bibinfo{author}{Danilatos, G.}
\newblock \bibinfo{title}{Review and outline of environmental sem at present}.
\newblock \emph{\bibinfo{journal}{J. Microsc.}} \textbf{\bibinfo{volume}{162}}, \bibinfo{pages}{391--402} (\bibinfo{year}{1991}) .

\bibitem{danilatos1988foundations}
\bibinfo{author}{Danilatos, G.}
\newblock \bibinfo{title}{Foundations of environmental scanning electron microscopy}.
\newblock \emph{\bibinfo{journal}{Advances in Electronics and Electron Physics}} \textbf{\bibinfo{volume}{71}}, \bibinfo{pages}{109--250} (\bibinfo{year}{1988}) .

\bibitem{assa2018monte}
\bibinfo{author}{Assa’d, A.}
\newblock \bibinfo{title}{Monte carlo calculation of the backscattering coefficient of thin films of low on high atomic number materials and the reverse as a function of the incident electron energy and film thickness}.
\newblock \emph{\bibinfo{journal}{Appl. Phys. A}} \textbf{\bibinfo{volume}{124}}, \bibinfo{pages}{699} (\bibinfo{year}{2018}) .

\bibitem{garcia2004momentum}
\bibinfo{author}{Garc{\'\i}a~de Abajo, F.~J.}
\newblock \bibinfo{title}{Momentum transfer to small particles by passing electron beams}.
\newblock \emph{\bibinfo{journal}{Phys. Rev. B}} \textbf{\bibinfo{volume}{70}}, \bibinfo{pages}{115422} (\bibinfo{year}{2004}) .

\bibitem{mifsud2022comparative}
\bibinfo{author}{Mifsud, D.~V.} \emph{et~al.}
\newblock \bibinfo{title}{Comparative electron irradiations of amorphous and crystalline astrophysical ice analogues}.
\newblock \emph{\bibinfo{journal}{Phys. Chem. Chem. Phys.}} \textbf{\bibinfo{volume}{24}}, \bibinfo{pages}{10974--10984} (\bibinfo{year}{2022}) .

\bibitem{adrian1984cryo}
\bibinfo{author}{Adrian, M.}, \bibinfo{author}{Dubochet, J.}, \bibinfo{author}{Lepault, J.} \& \bibinfo{author}{McDowall, A.~W.}
\newblock \bibinfo{title}{Cryo-electron microscopy of viruses}.
\newblock \emph{\bibinfo{journal}{Nature}} \textbf{\bibinfo{volume}{308}}, \bibinfo{pages}{32--36} (\bibinfo{year}{1984}) .

\bibitem{hong2018three}
\bibinfo{author}{Hong, Y.} \emph{et~al.}
\newblock \bibinfo{title}{Three-dimensional in situ electron-beam lithography using water ice}.
\newblock \emph{\bibinfo{journal}{Nano Lett.}} \textbf{\bibinfo{volume}{18}}, \bibinfo{pages}{5036--5041} (\bibinfo{year}{2018}) .

\bibitem{abellan2023radiolysis}
\bibinfo{author}{Abellan, P.}, \bibinfo{author}{Gautron, E.} \& \bibinfo{author}{LaVerne, J.~A.}
\newblock \bibinfo{title}{Radiolysis of thin water ice in electron microscopy}.
\newblock \emph{\bibinfo{journal}{J. Phys. Chem. C}} \textbf{\bibinfo{volume}{127}}, \bibinfo{pages}{15336--15345} (\bibinfo{year}{2023}) .

\bibitem{xu2020electron}
\bibinfo{author}{Xu, H.}, \bibinfo{author}{{\AA}ngström, J.}, \bibinfo{author}{Eklund, T.} \& \bibinfo{author}{Amann-Winkel, K.}
\newblock \bibinfo{title}{Electron beam-induced transformation in high-density amorphous ices}.
\newblock \emph{\bibinfo{journal}{J. Phys. Chem. B}} \textbf{\bibinfo{volume}{124}}, \bibinfo{pages}{9283--9288} (\bibinfo{year}{2020}) .

\bibitem{wu2021lithographic}
\bibinfo{author}{Wu, S.}, \bibinfo{author}{Zhao, D.}, \bibinfo{author}{Yao, G.}, \bibinfo{author}{Hong, Y.} \& \bibinfo{author}{Qiu, M.}
\newblock \bibinfo{title}{Lithographic properties of amorphous solid water upon exposure to electrons}.
\newblock \emph{\bibinfo{journal}{Appl. Surf. Sci.}} \textbf{\bibinfo{volume}{539}}, \bibinfo{pages}{148265} (\bibinfo{year}{2021}) .

\bibitem{han2022three}
\bibinfo{author}{Han, F.} \emph{et~al.}
\newblock \bibinfo{title}{Three-dimensional nanofabrication via ultrafast laser patterning and kinetically regulated material assembly}.
\newblock \emph{\bibinfo{journal}{Science}} \textbf{\bibinfo{volume}{378}}, \bibinfo{pages}{1325--1331} (\bibinfo{year}{2022}) .

\bibitem{obata2013high}
\bibinfo{author}{Obata, K.}, \bibinfo{author}{El-Tamer, A.}, \bibinfo{author}{Koch, L.}, \bibinfo{author}{Hinze, U.} \& \bibinfo{author}{Chichkov, B.~N.}
\newblock \bibinfo{title}{High-aspect 3d two-photon polymerization structuring with widened objective working range (wow-2pp)}.
\newblock \emph{\bibinfo{journal}{Light Sci. Appl.}} \textbf{\bibinfo{volume}{2}}, \bibinfo{pages}{e116--e116} (\bibinfo{year}{2013}) .

\bibitem{gu20253d}
\bibinfo{author}{Gu, S.} \emph{et~al.}
\newblock \bibinfo{title}{3d nanolithography with metalens arrays and spatially adaptive illumination}.
\newblock \emph{\bibinfo{journal}{Nature}} \textbf{\bibinfo{volume}{648}}, \bibinfo{pages}{591--599} (\bibinfo{year}{2025}) .

\bibitem{mastrangelo2002mechanical}
\bibinfo{author}{Mastrangelo, C.} \& \bibinfo{author}{Hsu, C.}
\newblock \bibinfo{title}{Mechanical stability and adhesion of microstructures under capillary forces. ii. experiments}.
\newblock \emph{\bibinfo{journal}{J. Microelectromech. Syst.}} \textbf{\bibinfo{volume}{2}}, \bibinfo{pages}{44--55} (\bibinfo{year}{2002}) .

\bibitem{namatsu1995dimensional}
\bibinfo{author}{Namatsu, H.}, \bibinfo{author}{Kurihara, K.}, \bibinfo{author}{Nagase, M.}, \bibinfo{author}{Iwadate, K.} \& \bibinfo{author}{Murase, K.}
\newblock \bibinfo{title}{Dimensional limitations of silicon nanolines resulting from pattern distortion due to surface tension of rinse water}.
\newblock \emph{\bibinfo{journal}{Appl. Phys. Lett.}} \textbf{\bibinfo{volume}{66}}, \bibinfo{pages}{2655--2657} (\bibinfo{year}{1995}) .

\bibitem{teufel2011sideband}
\bibinfo{author}{Teufel, J.~D.} \emph{et~al.}
\newblock \bibinfo{title}{Sideband cooling of micromechanical motion to the quantum ground state}.
\newblock \emph{\bibinfo{journal}{Nature}} \textbf{\bibinfo{volume}{475}}, \bibinfo{pages}{359--363} (\bibinfo{year}{2011}) .

\bibitem{li2021recent}
\bibinfo{author}{Li, P.} \emph{et~al.}
\newblock \bibinfo{title}{Recent advances in focused ion beam nanofabrication for nanostructures and devices: Fundamentals and applications}.
\newblock \emph{\bibinfo{journal}{Nanoscale}} \textbf{\bibinfo{volume}{13}}, \bibinfo{pages}{1529--1565} (\bibinfo{year}{2021}) .

\bibitem{huber2022tailoring}
\bibinfo{author}{Huber, R.} \emph{et~al.}
\newblock \bibinfo{title}{Tailoring electron beams with high-frequency self-assembled magnetic charged particle micro optics}.
\newblock \emph{\bibinfo{journal}{Nat. Commun.}} \textbf{\bibinfo{volume}{13}}, \bibinfo{pages}{3220} (\bibinfo{year}{2022}) .

\bibitem{shang2023dual}
\bibinfo{author}{Shang, X.}, \bibinfo{author}{Wang, N.}, \bibinfo{author}{Zhou, N.} \& \bibinfo{author}{Qiu, M.}
\newblock \bibinfo{title}{A dual-axis mechanical model for analyzing the capillary-force-induced clustering on periodic structures}.
\newblock \emph{\bibinfo{journal}{J. Appl. Phys.}} \textbf{\bibinfo{volume}{134}} (\bibinfo{year}{2023}) .

\end{thebibliography}
\end{flushleft}

\section{Methods}

\subsection{In situ environmental control and synergistic thermal management}
\label{8.1}
Fig.~\ref{Fig1}b presents an overall schematic of the three components of the FINEST platform. Inside the chamber, the operating temperature of the probe is tunable from -50~°C to 50~°C, and the system operates in low-vacuum and ESEM modes within a pressure range of 10~Pa to 2600~Pa. During operation, a tunable laser sweeps across the wavelength range of 1500~nm to 1630~nm. The F-P cavity interference signal, which carries the force information applied to the probe, is continuously recorded by a power meter in real time.

To ensure the FINEST platform operates reliably under e-beam irradiation and yields accurate mechanical measurements, thermal effects represent a critical challenge that must be addressed. Conventional SEMs operate in a high vacuum environment (typically $< 10^{-2}$~Pa), which renders thermal convection largely ineffective. Consequently, the polymeric resin suffers from thermal ablation under mW-level laser irradiation (Fig.~\ref{Fig1}c, left). 

Therefore, to ensure the proper functioning of the probe, we implemented a threefold synergistic optimization strategy: (1) transitioning from standard SEM to ESEM mode (which increases the chamber pressure while preserving e-beam focusing and bombardment capabilities), (2) introducing active cooling (maintaining the temperature below -10~°C), and (3) coating the probe with an Ag film to enhance thermal conductivity. Following these optimizations, after 10~h of continuous operation under 20~mW laser irradiation and persistent e-beam bombardment, the readout signal of the FINEST remained stable and identical to the initial measurement. Furthermore, SEM characterization confirmed that the surface details of the probe exhibited no observable degradation (see Fig.~\ref{Fig1}c). This demonstrates that our synergistic strategy successfully equips the probe with exceptional adaptability to extreme environments, enabling it to maintain a high degree of geometrical and mechanical consistency over extended periods.

\subsection{Finite-element simulation of the FINEST probe}
\label{8.2}
Finite-element simulations were performed using COMSOL Multiphysics 6.1 to evaluate the mechanical performance of the R100 FINEST probe. The simulated three-dimensional stress and displacement fields are presented in Fig.~\ref{Fig1}d. A multiphysics coupling architecture combining the Solid Mechanics and Shell interfaces was employed. The three-dimensional micro-helical geometry was imported from SolidWorks. The geometric model strictly replicated the actual fabricated dimensions. Furthermore, the cutting features generated during the FIB process were accurately incorporated. Specifically, the residual support pillars were explicitly reconstructed. Each pillar was divided into three segments, with a $10~\mu\mathrm{m}$ section removed from the center of each segment.

The structural material for the main body was defined as a negative-tone photoresist, possessing a density of $1200~\mathrm{kg/m^{3}}$, a Young's modulus of $3~\mathrm{GPa}$, and a Poisson's ratio of $0.35$. Excluding the bottom surface fixed to the optical fiber, all outer surfaces of the helical structure were assigned a silver film with a thickness of $23.3~\mathrm{nm}$. This metallic layer was modeled via the Shell interface using standard silver properties extracted from the MEMS material library, featuring a Young's modulus of $83~\mathrm{GPa}$, a density of $10500~\mathrm{kg/m^{3}}$, and a Poisson's ratio of $0.37$. A Solid-Thin Structure Connection multiphysics coupling node was implemented to achieve synergistic stress analysis between the three-dimensional micro-helical main body and the thin film.

A fixed constraint was applied to the annular base of the probe. All remaining external surfaces were designated as free boundaries. A stationary study utilizing a steady-state solver was executed to compute the mechanical response under a predefined boundary load applied to the top plate. The load was set to $1~\mathrm{pN}$ directed vertically downward. A tetrahedral mesh was generated, with refined elements along the helical arms and support pillars to accurately resolve local deformations.

To ensure numerical accuracy, a systematic mesh convergence analysis was performed. Using the displacement magnitude of the three-dimensional cut point at the center of the top plate as the core parameter, grid independence was confirmed when the relative change in displacement caused by successive mesh refinements dropped below $1\%$. Simultaneously, the relative residual of the steady-state solver strictly converged below a tolerance threshold of $10^{-3}$, ensuring the stability of the static numerical iterations. Following the simulation, the three-dimensional stress distribution and displacement fields were extracted. Additionally, the spatial displacement of the three-dimensional cut point located at the center of the top plate was evaluated. This specific displacement magnitude was directly acquired through a one-dimensional point evaluation plot group, thereby enabling the simulated calculation of the spring constant $k$.

\subsection{Fabrication optimization and structural characterization}
\label{8.3}
To achieve femtonewton-level detection sensitivity, the helical probe must possess an extremely low spring constant. However, this extreme mechanical flexibility fundamentally conflicts with the structural stability required during conventional micro/nanofabrication. Direct application of standard fabrication workflows inevitably leads to severe structural failure, including localized polymer bubbling from uniform optical exposure, capillary-force-induced structural collapse during liquid-phase development\cite{shang2023dual}, and bending or fracture caused by asymmetric stress release during focused ion beam (FIB) milling (Extended Data Fig.~\ref{Ex1}a). A comprehensive synergistic optimization strategy was therefore developed to overcome this fabrication bottleneck.

First, sacrificial scaffold structures were integrated into the probe design to provide essential rigid support during the initial processing steps (Fig.~\ref{Fig2}a,b and Extended Data Fig.~\ref{Ex1}a). In addition, a spatially modulated scanning speed was employed during the two-photon lithography stage to suppress localized thermal damage. Standard isopropanol was replaced with a low-surface-tension fluorinated solvent ($\mathrm{C_4F_9OCH_3}$) during development, which significantly reduced capillary forces at the liquid--gas interface. As a result, the fragile helical structure was protected from surface-tension-induced collapse. Finally, strict vertical alignment and clamping were implemented during scaffold removal by FIB milling, ensuring symmetric detachment of the support beams (Fig.~\ref{Fig2}a and Extended Data Fig.~\ref{Ex1}a). Collectively, these process optimizations overcome the yield limitations associated with fabricating ultra-flexible three-dimensional microstructures and establish the basis for high-fidelity manufacturing of the FINEST probe.

Step-by-step characterization of the probe throughout the fabrication process verifies the reliability of this architecture-inspired manufacturing strategy. During the initial two-photon lithography stage, \textit{in situ} optical monitoring ensured precise coaxial alignment between the three-dimensional direct-writing field of view and the single-mode fiber core (Extended Data Fig.~\ref{Ex1}b). This alignment is the primary optical prerequisite for constructing an efficient fiber-tip Fabry--P\'{e}rot cavity. Following low-surface-tension development and ultraviolet curing, three-dimensional digital microscopy and high-resolution helium ion microscopy confirmed that the polymer skeleton maintained high structural fidelity after removal from the liquid environment (Extended Data Fig.~\ref{Ex1}c,d). Supported by the scaffold structure, the ultra-flexible helical beams and suspended top plate avoided capillary-force-induced distortion and collapse, retaining their designed geometric symmetry (Fig.~\ref{Fig2}d,e and Extended Data Fig.~\ref{Ex1}c,d).

Subsequently, a conformal metallization process using magnetron sputtering uniformly deposited an Ag film across the structural surface (Extended Data Fig.~\ref{Ex1}e). This metal layer provided the optical reflectance required for Fabry--P\'{e}rot interferometric readout and also served as a conductive and thermally dissipative interface under continuous electron-beam bombardment. Finally, the sacrificial scaffolds were precisely removed by FIB milling, releasing the fully suspended helical probe structure (Fig.~\ref{Fig2}f--h).

\subsection{Spectral preprocessing and sub-grid peak-finding algorithm}
\label{8.4}
Reliable extraction of extremely small resonance wavelength shifts induced by femtonewton-scale forces from discrete spectral sampling data requires specialized processing. To achieve this, a spectral peak-finding algorithm based on sub-grid analytical fitting was implemented. The physical scanning step of the tunable laser (e.g., $0.001$~nm) is relatively large. Consequently, directly extracting discrete extrema introduces truncation errors. During data preprocessing, a global Savitzky-Golay filter was initially applied to the raw interference spectrum. This filter utilizes a cubic polynomial. It effectively suppresses high-frequency optoelectronic measurement noise. Simultaneously, it strictly preserves the inherent envelope shape and spectral linewidth of the Fabry-Pérot interference fringes. Subsequently, spectral extrema are roughly located by setting a strict topological prominence threshold. This step filters out pseudo-peaks caused by local noise. An extremely narrow local window is then defined around these approximate peak locations. Quadratic polynomial regression is performed on the data points within this localized region. By calculating the analytical vertex of this locally fitted parabola, the algorithm successfully overcomes the resolution limits of the discrete hardware sampling grid. The previously demonstrated baseline stability relies directly on this high-precision analytical peak-finding strategy. This numerical analysis workflow eliminates reading uncertainties caused by the finite wavelength step. It provides the essential algorithmic foundation required for the system to approach its theoretical ultimate sensing performance.

\subsection{Monte Carlo simulation of electron energy deposition}
\label{8.5}
To quantitatively analyze the interaction processes, energy deposition distributions, and electron transport dynamics between the e-beam and the probe material, Monte Carlo simulations were conducted using the CASINO V2.5.1 software. In the geometric model, the probe was equivalently configured as a bilayer structure. The top layer consists of a $23.30~\mathrm{nm}$ thick silver (Ag) conductive film with a density set to $10500~\mathrm{kg/m^3}$; the bottom layer is a $2500~\mathrm{nm}$ thick negative photoresist layer with a density of $1200~\mathrm{kg/m^3}$. The e-beam radius was set to $10~\mathrm{nm}$, and the sample tilt angle was $3^\circ$. To systematically evaluate the impact of different acceleration voltages on electron penetration depth and momentum transfer, incident electron energies of $1~\mathrm{keV}$, $10~\mathrm{keV}$, and $20~\mathrm{keV}$ were selected for the simulations, with the total number of incident electrons set to $20,000$ for each independent run.

Regarding the physical mechanisms, the simulation employed an interpolation based Mott cross-section model to calculate the total and partial elastic scattering cross sections of electrons. For the evaluation of inelastic scattering energy loss ($\text{d}E/\text{d}S$) and ionization potential within the material, the empirical model proposed by Joy and Luo (1989) was utilized. The effective ionization cross section was calculated based on the Casnati model, whereas the direction cosines of the electron after each collision scattering event were solved using the Drouin (1996) model. The Monte Carlo random sampling process adopted the random number generation algorithm by Press et al. (1986). The cutoff energy threshold for tracking individual electron trajectories was set to $0.05~\mathrm{keV}$; electrons falling below this threshold were considered to have their energy fully deposited within the local material volume.

In the data post processing stage, the system extracted the two dimensional projection of energy deposition on the $XZ$ plane within a three dimensional space (divided into a $50 \times 50 \times 50$ grid) and integrated the energy values along the projection axis. Concurrently, constant energy contour profiles were generated via a data interpolation algorithm to precisely delineate the energy deposition gradients at varying penetration depths. Furthermore, the energy and angular distributions of both backscattered and transmitted electrons were comprehensively recorded.

\subsection{\textit{In-situ} ice deposition and e-beam interactions}
\label{8.6}
The core system for ice deposition and etching comprises a customized scanning electron microscope (ZEISS Sigma 300) highly integrated with a high-precision pattern generator (Raith Elphy Quantum). Polished silicon wafers with native oxide layers were utilized as experimental substrates. The extreme low-temperature states of the system were continuously monitored using a sensory network constructed from platinum resistance thermometers and a data acquisition multimeter (Model 2700, Keithley, US). During the preparation of the gas-phase precursor, magnesium sulfate heptahydrate crystals were sublimated in a vacuum. The resulting vapor was strictly filtered through a microfiltration membrane with a pore size of $0.22~\mu\mathrm{m}$ to ensure a stable output of high-purity water molecules.

To guarantee the absolute thickness uniformity of the large-area ice layer, the sample stage was precisely positioned approximately $3~\mathrm{mm}$ below the gas injection nozzle. Prior to ice deposition, the electron gun valve of the microscope was strictly closed to completely prevent the reverse diffusion of water molecules into the electron optical system. Subsequently, a precision leak valve (VAT, Switzerland) was opened. Uniform area deposition of amorphous ice was achieved under the continuous monitoring of the pressure drop within the water vapor chamber using a capacitance vacuum gauge (INFICON, Switzerland).

During the patterning and etching phase, the area exposure mode was exclusively employed. Because the amorphous ice layer is extremely thin, conventional focusing procedures were challenging to implement. A single-point e-beam was utilized to instantaneously ablate a nanoscale physical circular hole in the ice layer outside the target processing area. By serving as an absolute reference benchmark, this feature enabled the precise locking of the ultimate focal plane and the elimination of astigmatism. Ultimately, the pattern generator converted the geometric array data into voltage signals for the deflection coils, driving the e-beam to complete the entire direct-writing process according to the predefined area exposure dose.

\section*{Data availability}

All data are available in the main text, Methods or in the Supplementary Information. Other information related to this study is available from the corresponding author on request.

\section*{Acknowledgements}
This work was supported by the National Natural Science Foundation of China (grant nos. U25A203659 and 61927820). We thank Wei Yan (Qiu Min Laboratory, Westlake University) for his contributions to the theoretical aspects of this work, and Ruofei Chen (Zhejiang University) for insightful discussions and valuable perspectives from the field of architecture. We also acknowledge the Westlake Center for Micro/Nano Fabrication for facility support and technical assistance.

\section*{Author contributions}

C.L., X.G.S. and M.Q. conceived the study and designed the overall experiments. M.Q. supervised the project. C.L. modelled, fabricated and characterized the FINEST probe. C.L. and X.Y.S. built the optical radiation platform and performed the calibration experiments. C.L., X.L. and W.C. designed the experimental protocol for e-beam measurements, built the experimental platform, and performed measurements. C.L., K.Z. and X.W. designed and performed the amorphous-ice etching experiments. Y.Y. modelled and simulated the reflectance of structures. C.L., X.L. and M.Q. developed the model for e-beam force and electron--matter interactions. C.L. developed the capillary-force model and performed the Monte Carlo simulations. C.L. prepared the manuscript, with revisions from all authors.

\section*{Competing interests}

Chinese patent applications related to this work have been filed, with M.Q., X.G.S., and C.L. as co-inventors. The authors declare no other competing interests.

\renewcommand{\figurename}{Extended Data Fig.}
\setcounter{figure}{0}
\renewcommand{\theHfigure}{ED\arabic{figure}}

\clearpage
\begin{figure}[htbp]
\centering
\includegraphics[width=0.9\textwidth]{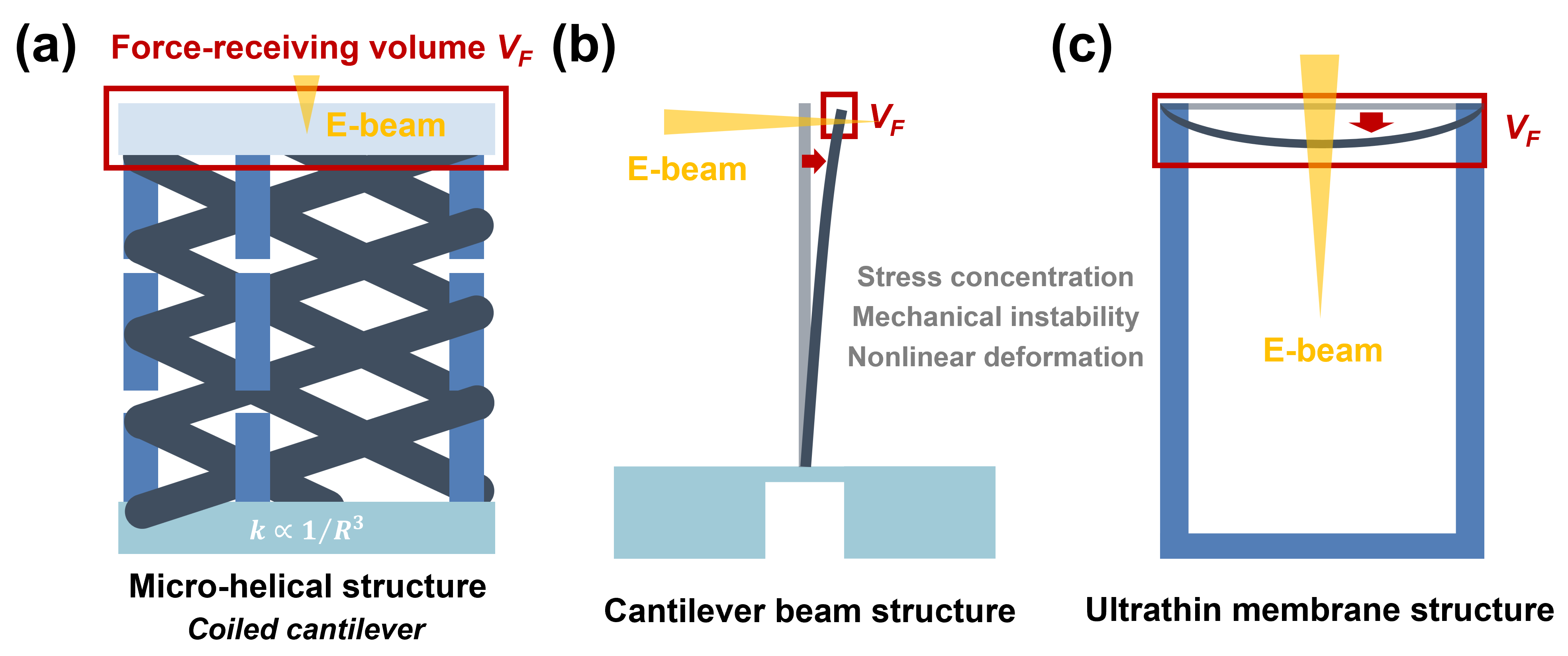}
\caption{\textbf{Schematic comparison of nanomechanical sensing architectures.} 
\textbf{(a)} The three-dimensional micro-helical structure under e-beam incidence. The red box indicates the expanded force-receiving volume ($V_F$) defined by the flat circular top plate. 
\textbf{(b)} A conventional cantilever beam structure under e-beam incidence. The schematic highlights a laterally restricted $V_F$ dictated by the slender geometry of the lever. 
\textbf{(c)} An ultrathin membrane structure under e-beam incidence, illustrating a longitudinally restricted $V_F$ bounded by the membrane thickness that leads to direct e-beam penetration. Red arrows indicate the direction of structural deformation, with faded geometries representing the undeformed states. Both traditional configurations (\textbf{b} and \textbf{c}) are inherently susceptible to structural vulnerabilities, including stress concentration, mechanical instability, and nonlinear deformation.}
\label{Ex0}
\end{figure}

\clearpage
\begin{figure}[htbp]
\centering
\includegraphics[width=0.9\textwidth]{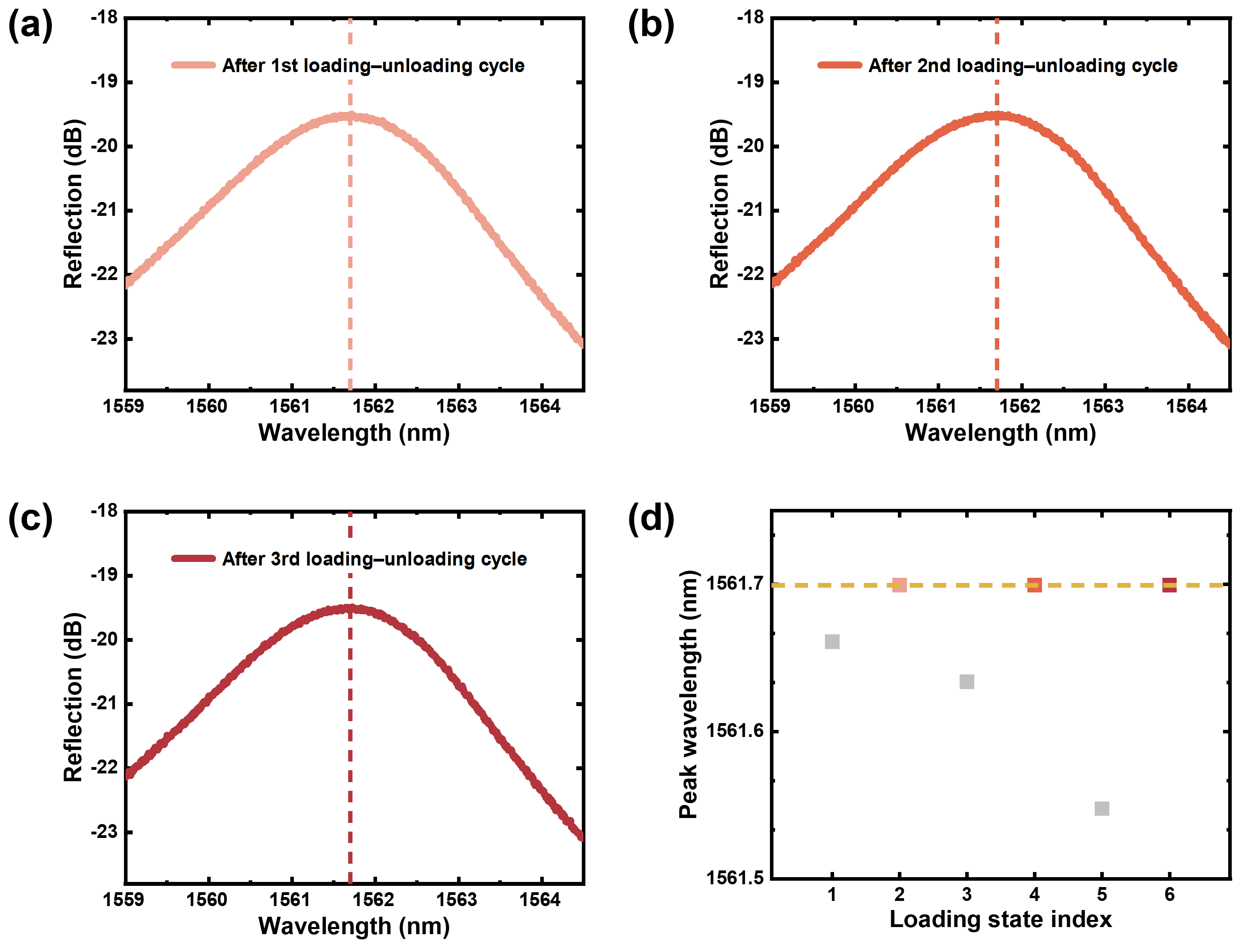}
\caption{\textbf{Baseline stability and performance metrics of the FINEST probe.} 
\textbf{(a)--(c)} Reflection spectra recorded after the 1st, 2nd, and 3rd electron-beam loading--unloading cycles, respectively. 
\textbf{(d)} Peak wavelength variation across successive loading--unloading cycles. Grey and colored squares denote the loaded (electron-beam on) and unloaded (static) states, respectively, demonstrating a stable baseline (dashed line).}
\label{Ex5}
\end{figure}

\clearpage
\begin{figure}[htbp]
\centering
\includegraphics[width=1\textwidth]{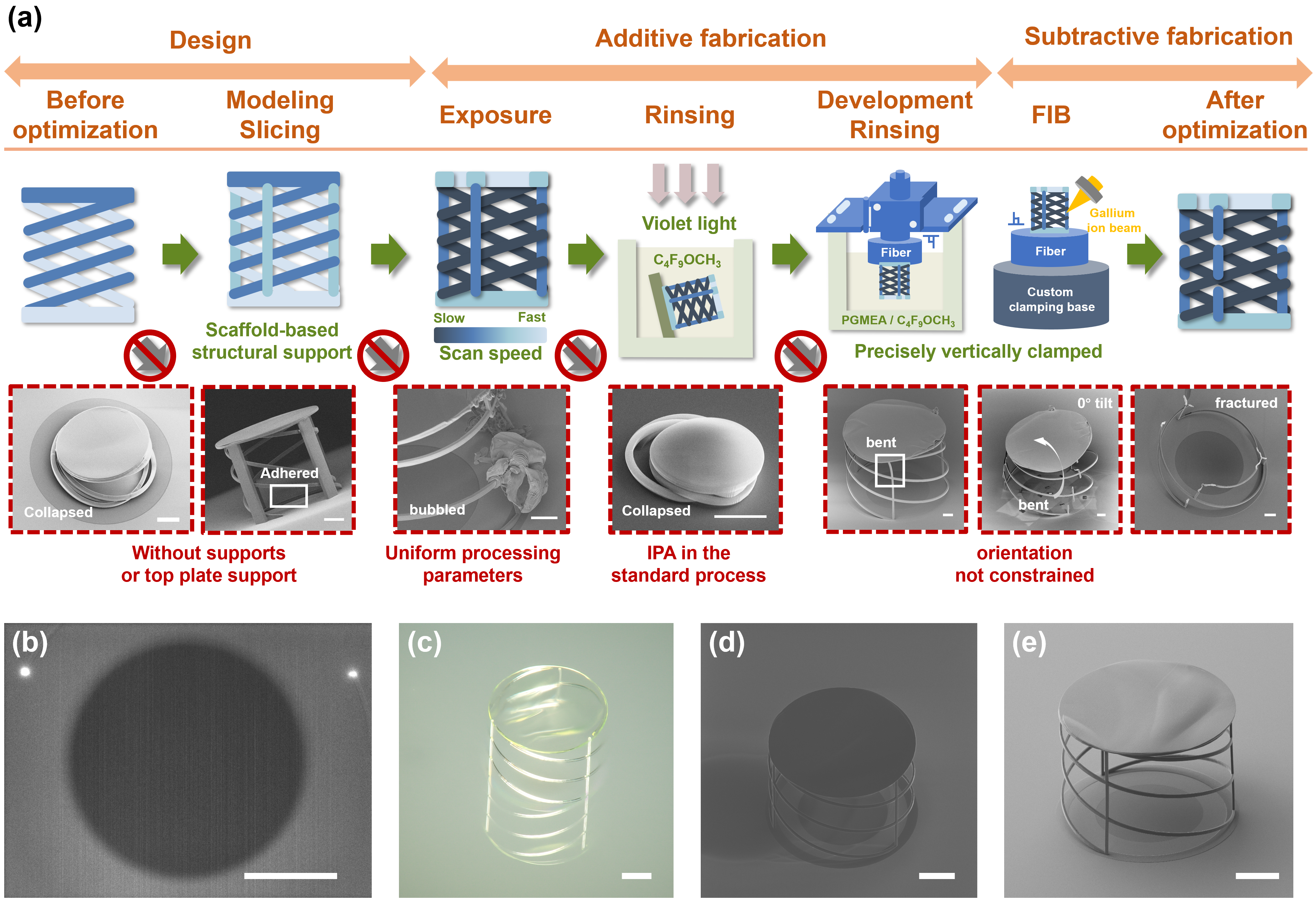}
\caption{\textbf{Optimization workflow and stepwise structural characterization of the FINEST probe.} 
\textbf{(a)} Schematic of the synergistic optimization strategy for fabricating the FINEST probe. The top panel illustrates the optimized pipeline, highlighting the introduction of sacrificial scaffolds, spatially modulated scan speeds, low-surface-tension fluorinated solvent (\ce{C4F9OCH3}) rinsing, and precisely vertically clamped post-processing. Bottom insets (red dashed boxes) show SEM/FIB images of typical catastrophic failures. 
\textbf{(b)} \textit{In situ} optical micrograph acquired during the TPL stage. 
\textbf{(c)} 3D digital microscope image and \textbf{(d)} HIM image of the cross-linked polymer skeleton after development. 
\textbf{(e)} SEM image after conformal metallization. Scale bars, 20~$\mu$m (\textbf{a}, insets) and 50~$\mu$m (\textbf{b}--\textbf{e}).}
\label{Ex1}
\end{figure}

\clearpage
\begin{figure}[htbp]
\centering
\includegraphics[width=0.6\textwidth]{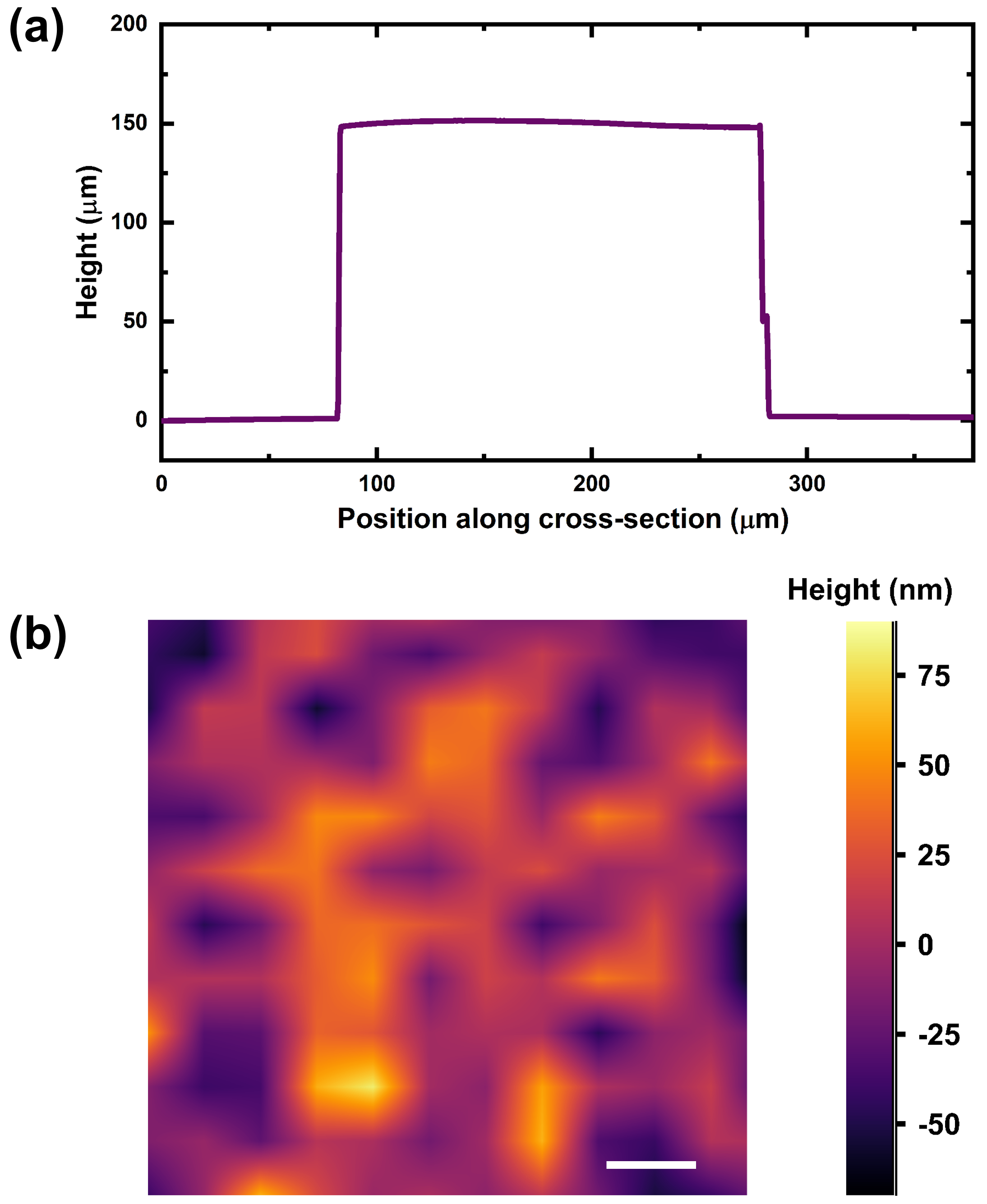}
\caption{\textbf{Confocal microscopic topographic characterization of the FINEST probe.} 
\textbf{(a)} Cross-sectional height profile extracted across the center of the top plate. 
\textbf{(b)} Local high-resolution topographic map of the top plate surface. Scale bar, 1~$\mu$m.}
\label{Ex6}
\end{figure}

\clearpage
\begin{figure}[htbp]
\centering
\includegraphics[width=1\textwidth]{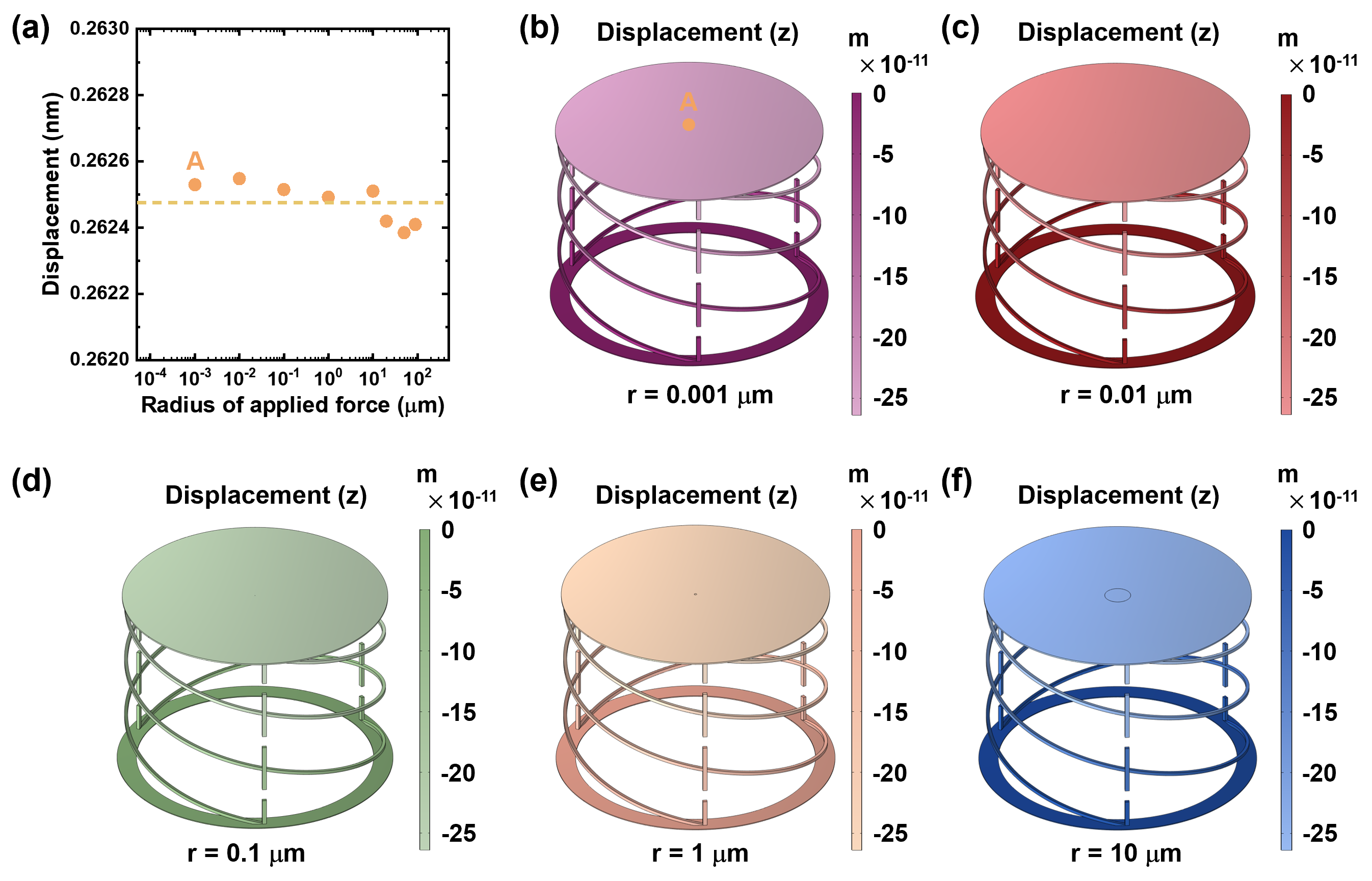}
\caption{\textbf{Numerical validation of the scale-independent mechanical response of the FINEST probe.} 
\textbf{(a)} Simulated vertical displacement of the FINEST probe under a constant 1~pN load as a function of the force radius. 
\textbf{(b)--(f)} FEA results showing the 3D displacement fields of the probe under localized loading with radii of 0.001~$\mu$m, 0.01~$\mu$m, 0.1~$\mu$m, 1~$\mu$m, and 10~$\mu$m, respectively.}
\label{Ex4}
\end{figure}

\clearpage
\begin{figure}[htbp]
\centering
\includegraphics[width=0.6\textwidth]{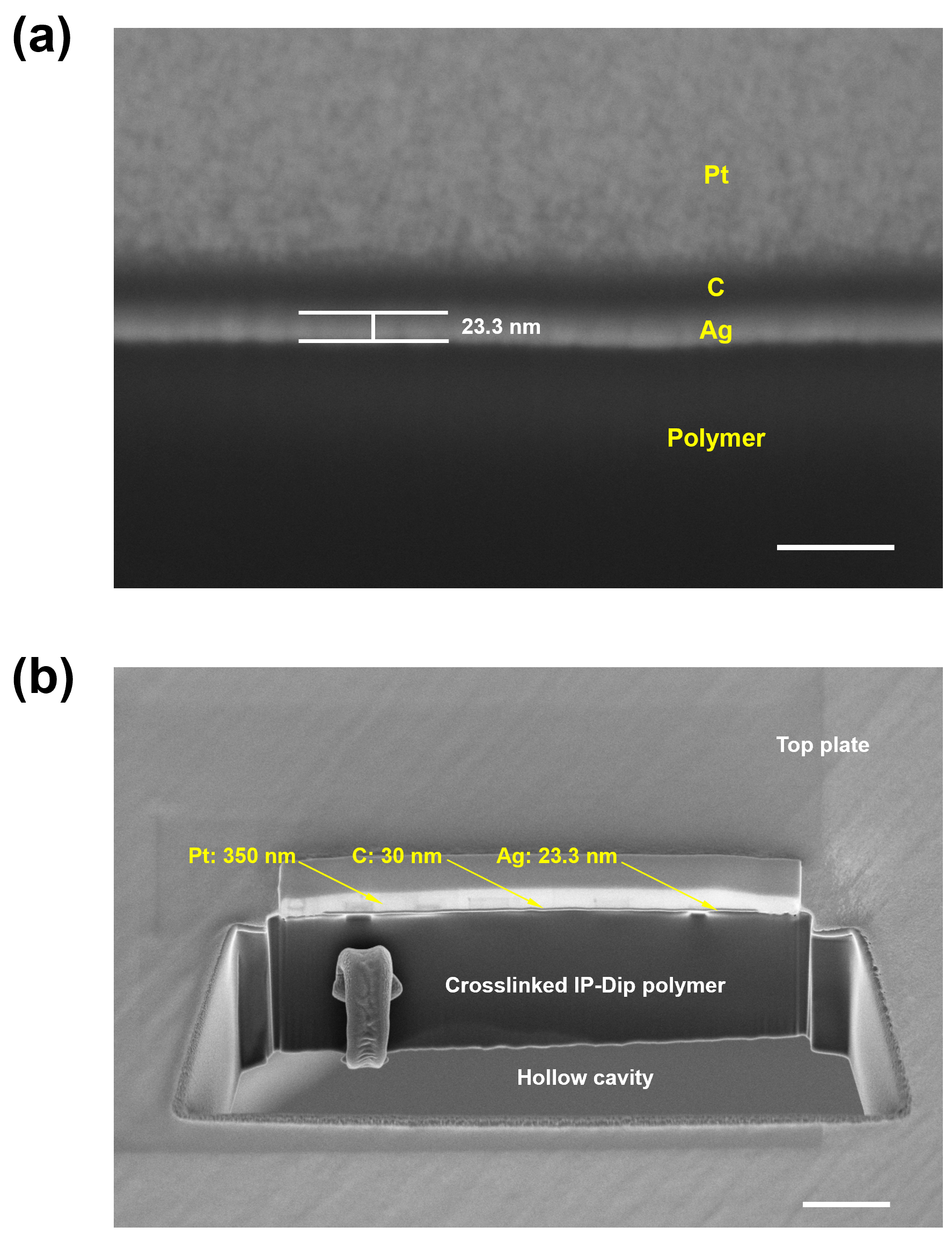}
\caption{\textbf{Cross-sectional characterization of the Ag thin film on the probe top plate.} 
\textbf{(a)} SEM image of the heterostructure interface, resolving the 23.3~nm Ag layer on the cross-linked polymer substrate. Scale bars, 100~nm.
\textbf{(b)} Low-magnification SEM image showing the FIB-milled cross-section of the top plate. Scale bars, 2~$\mu$m.}
\label{Ex2}
\end{figure}

\clearpage
\begin{figure}[htbp]
\centering
\includegraphics[width=1\textwidth]{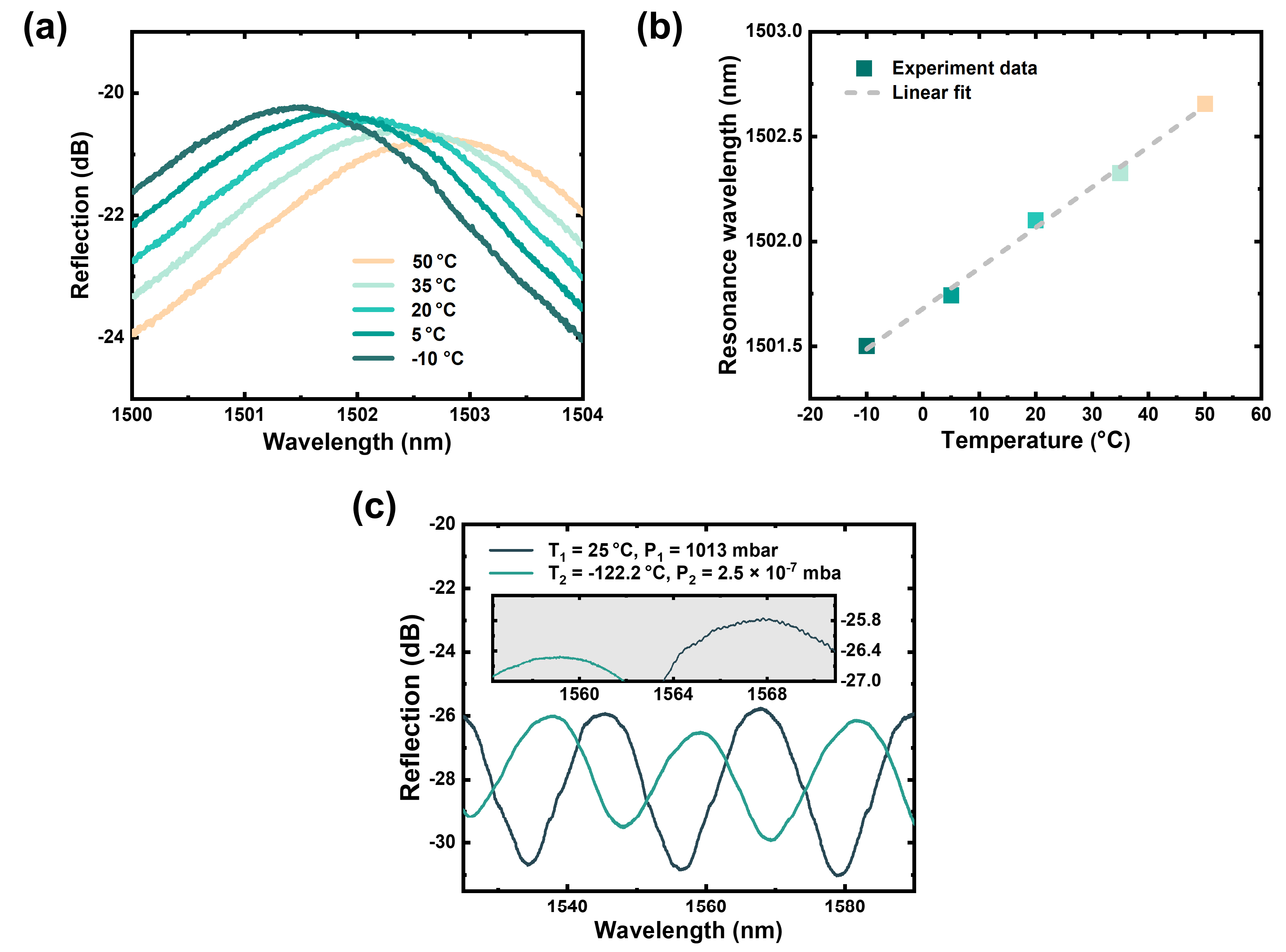}
\caption{\textbf{Thermal stability and temperature cross-sensitivity of the FINEST probe.} 
\textbf{(a)} Evolution of the Fabry--P\'{e}rot resonance spectra recorded across a temperature range from $-10$~$^\circ$C to 50~$^\circ$C. 
\textbf{(b)} Linear regression analysis of the resonance wavelength shift as a function of temperature. The extracted temperature cross-sensitivity is 19.1~pm/$^\circ$C.
\textbf{(c)} Interference resonance spectra of the FINEST probe under static conditions at different environmental parameters. Compared with ambient-pressure and room-temperature conditions, the spectrum acquired under low-temperature and low-pressure conditions is smoother and exhibits a higher signal-to-noise ratio. Inset, zoomed-in view.}
\label{Ex3}
\end{figure}

\clearpage
\begin{figure}[htbp]
\centering
\includegraphics[width=1\textwidth]{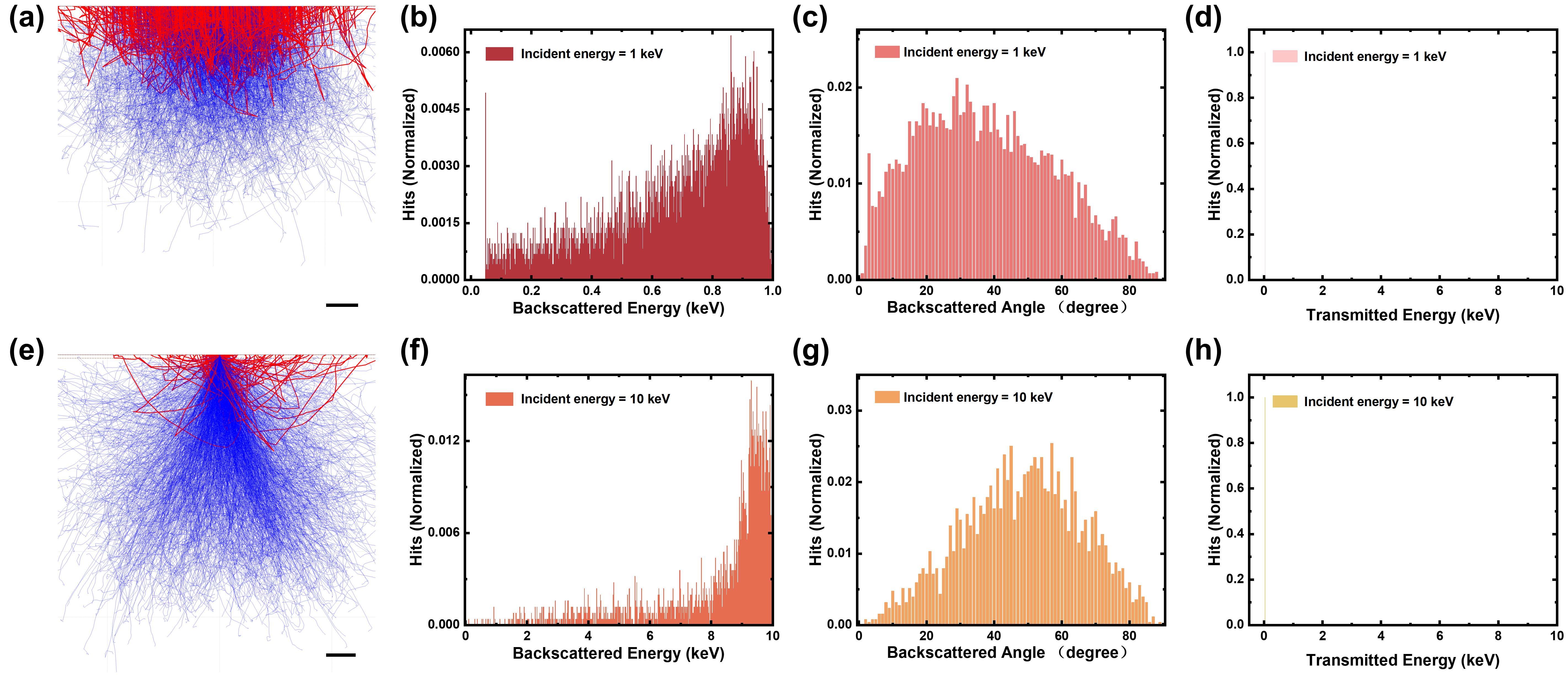}
\caption{\textbf{E-beam transport dynamics and statistical distributions at low-to-medium incident energies.} 
  \textbf{(a, e)} Monte Carlo simulation of electron trajectories within the probe material at EHTs of 1~keV and 10~keV. Red and blue tracks represent backscattered and trapped electrons, respectively. 
    \textbf{(b, f)} Histograms of backscattered energy for 1~keV and 10~keV, showing broad continuous energy distributions near the initial incident energy. 
    \textbf{(c, g)} Angular distributions of backscattered electrons relative to the surface normal, indicating significant angular divergence induced by large-angle elastic scattering. 
    \textbf{(d, h)} Normalized transmitted energy spectra. The complete absence of transmitted hits at both 1~keV and 10~keV indicates that the penetration depth is insufficient to breach the probe thickness, thereby completely closing the transmission channel ($f_T = 0$).}
\label{Ex7}
\end{figure}

\clearpage
\begin{figure}[htbp]
\centering
\includegraphics[width=1\textwidth]{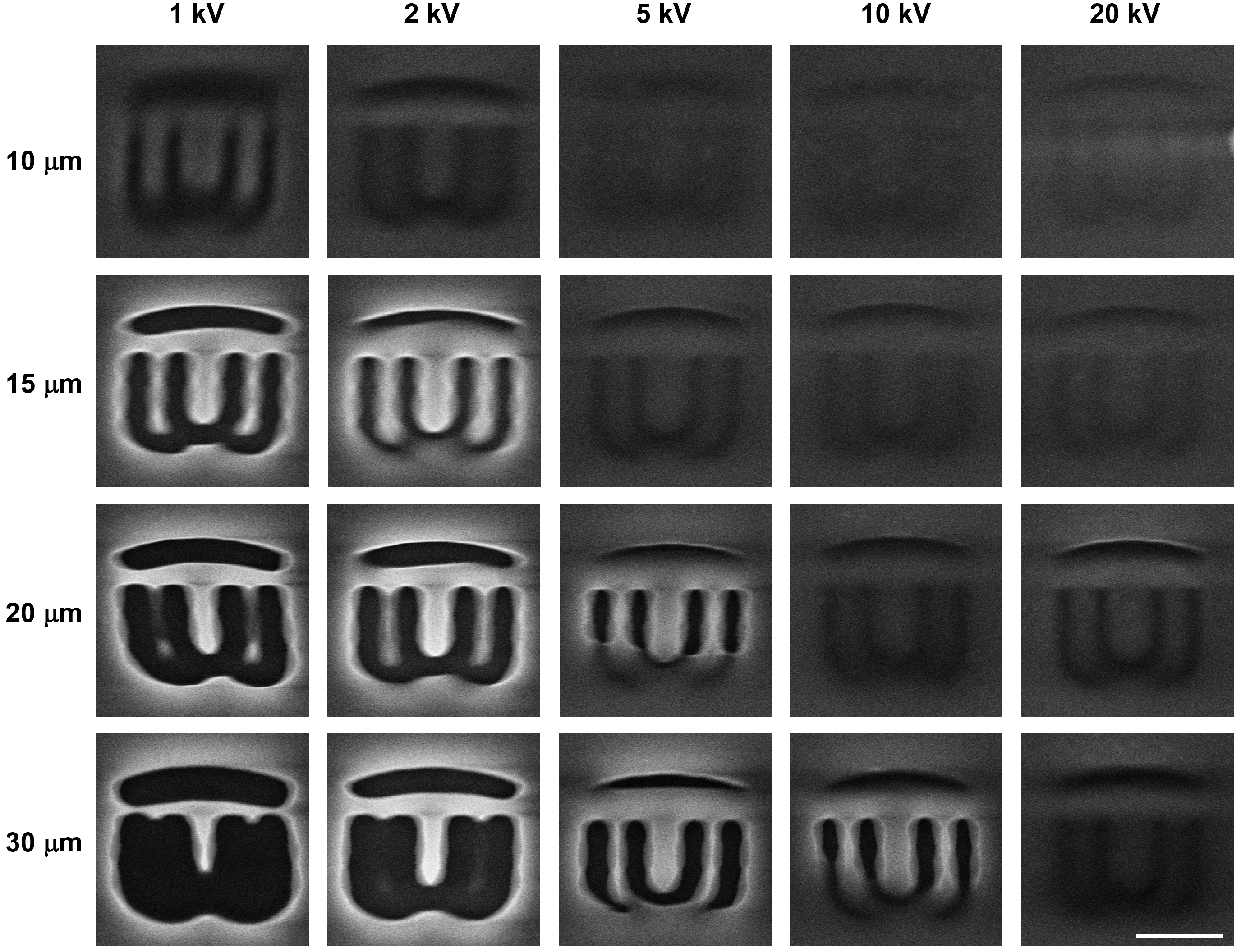}
\caption{\textbf{Morphological threshold of electron-beam-induced ice etching.} 
SEM images (0$^\circ$ tilt) of a 200~nm amorphous ice layer etched across a parameter matrix of EHTs (1--20~kV) and aperture sizes (10--30~$\mu$m). All images were acquired at a consistent imaging condition of EHT = 5~kV and a 15~mm aperture to ensure objective comparison of the etched morphologies. A diagonal boundary from top-left to bottom-right distinguishes the etching regimes: the region to the bottom-left represents complete removal of the ice layer, while the boundary itself marks the threshold of incipient breakthrough. Scale bar, 1~$\mu$m.}
\label{Ex8}
\end{figure}

\clearpage
\begin{figure}[htbp]
\centering
\includegraphics[width=0.8\textwidth]{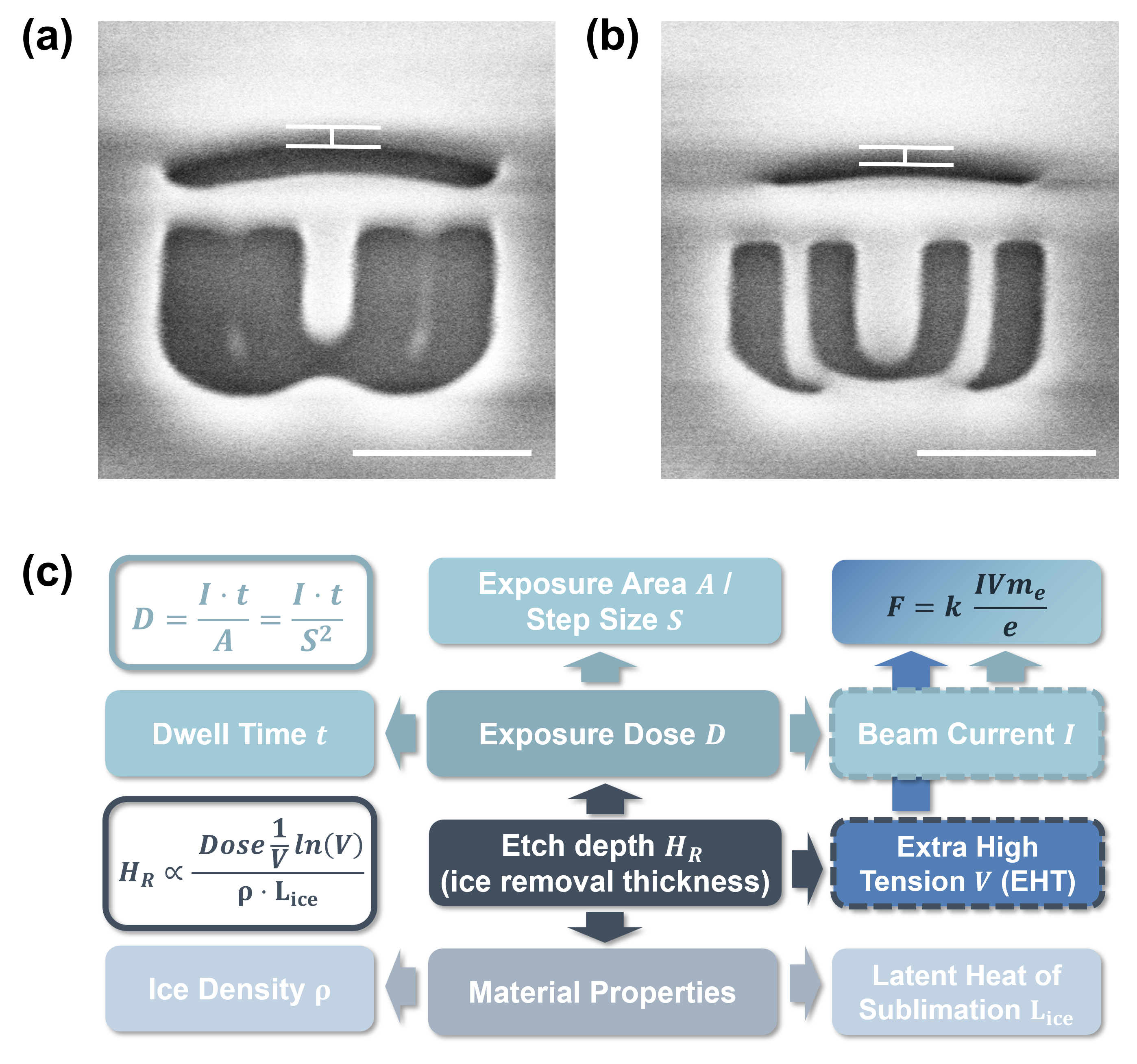}
\caption{\textbf{Thickness quantification and analytical modeling of e-beam ice etching.} 
\textbf{(a), (b)} Tilted SEM images (30$^\circ$) of etched amorphous ice patterns, facilitating the direct measurement of ice removal depth ($H_R$). The total initial thickness of the amorphous ice layer is 200~nm. Images were acquired at EHT = 5~kV with an aperture size of 30~$\mu$m. Scale bars, 1~$\mu$m.
\textbf{(c)} Framework assessing removal contributions. It correlates experimental control parameters---exposure dose ($D$), dwell time ($t$), beam current ($I$), and EHT ($V$)---with the measured etch depth ($H_R$) and the momentum-transfer force ($F$). This model provides a quantitative basis for comparing the observed ice removal against the theoretical contribution of mechanical momentum transfer.}
\label{Ex9}
\end{figure}


\end{document}